\documentclass[preprints, article, submit, pdftex, moreauthors]{Definitions/mdpi} 
\usepackage{graphicx}
\usepackage{subcaption}
\usepackage[nolist]{acronym}
\usepackage{dsfont}
\usepackage{amssymb}
\firstpage{1} 
\pubvolume{1}
\issuenum{1}
\articlenumber{0}
\pubyear{2024}
\copyrightyear{2024}
\datereceived{19.08.2024} 
\daterevised{ } 
\dateaccepted{ } 
\datepublished{ } 
\hreflink{https://doi.org/} 

\Title{Towards a Field Based Bayesian Evidence Inference from Nested Sampling Data}

\Author{Margret Westerkamp
 $^{1,2}$*
, Jakob Roth $^{1,2,3}$, Philipp Frank $^{1}$, Will
 Handley $^{4,5}$ and  Torsten Enßlin $^{1,2}$} 

\AuthorNames{Margret Westerkamp, Jakob Roth, Philipp Frank, Will
 Handley, Torsten Enßlin}

\AuthorCitation{Westerkamp, M.; Roth, J.; Frank, P. ; Handley, W.; Enßlin T.}

\address{%
$^1$ \quad Max Planck Institute for Astrophysics, 85748 Garching,
 Germany; roth@mpa-garching.mpg.de
 (J.R.); philipp@mpa-garching.mpg.de (P.F.); ensslin@mpa-garching.mpg.de (T.E.)
\\
$^2$ \quad Ludwig-Maximilians-Universität, Faculty for Physics, 80539
Munich, Germany
\\
$^3$ \quad Technical University Munich, Department for Computer Science,
 85748 Munich, Germany
\\
$^4$ \quad Cavendish Laboratory, Cambridge CB3 0HE, UK; wh260@cam.ac.uk (W.H.)
\\
$^5$ \quad Kavli Institute for Cosmology, Cambridge CB3 0EZ, UK
}
\corres{Correspondence: margret@mpa-garching.mpg.de}

\abstract{
Nested sampling (NS) is a stochastic method for computing the log-evidence of a Bayesian problem. It relies on stochastic estimates of prior volumes enclosed by likelihood contours, which limits the accuracy of the log-evidence calculation. We propose to transform the prior volume estimation into a Bayesian inference problem, which allows us to incorporate a smoothness assumption for likelihood-prior volume relations. As a result, we aim to increase the accuracy of the volume estimates and thus improve the overall log-evidence calculation using NS. The method presented works as a post-processing step for NS and provides posterior samples of the likelihood-prior-volume relation, from which the log-evidence can be calculated. We demonstrate an implementation of the algorithm and compare its results with plain NS on two synthetic datasets for which the underlying evidence is known. We find a significant improvement in accuracy for runs with less than one hundred active samples in NS, but are prone to numerical problems beyond this point.}


\begin{document}
\begin{acronym}
\acro{geoVI}[geoVI]{geometric variational inference}
\acro{LRPS}[LRPS]{likelihood restricted prior sampling}
\acro{LCDM}[LCDM]{concordance cosmological model }
\acro{KLCDM}[KLCDM]{concordance cosmological model with curvature}
\acro{IFT}[IFT]{information field theory}
\acro{KL}[KL]{Kullback-Leibler divergence}
\acro{MAP}[MAP]{maximum a posteriori}
\acro{MCMC}[MCMC]{Markov Chain Monte Carlo}
\acro{NIFTy}[NIFTy]{Numerical Information Field TheorY}
\acro{NS}[NS]{nested sampling}
\acro{NS-SMC}[NS-SMC]{nested sampling via sequential Monte Carlo}
\acro{VI}[VI]{variational inference}
\end{acronym}
\section{Introduction}
In Bayesian inference, we update our knowledge about a measure of interest, which we call the signal, $s$, on the basis of some given data, $d$. This signal can be, among other things, a set of model parameters $\theta_\mathcal{M}$, the model itself $\mathcal{M}$, or a continuous field $\varphi$. Given the prior $\mathcal{P}(s)$, which describes our a priori knowledge about $s \in \{ \theta_\mathcal{M}, \mathcal{M}, \varphi, ... \}$, the likelihood $\mathcal{P}(d|s)$ of the data given the signal and the normalising constant $\mathcal{P}(d) =  \int ds ~ \mathcal{P}(d|s) \mathcal{P}(s)$, known as the evidence, Bayes theorem returns the posterior of the signal given the data,
\begin{align}
\mathcal{P}(s|d) = \frac{\mathcal{P}(d|s) \mathcal{P}(s)}{\mathcal{P}(d)}. \label{eq:bayes}
\end{align}
In particular, Bayesian parameter estimation is concerned with updating knowledge about a set of model parameters given the data. Here, the model parameters in general depend on the given model. Accordingly, the quantity of interest is the posterior $\mathcal{P}(\theta_{\mathcal{M}}|d, \mathcal{M})$. 
In the case of field inference, we aim to reconstruct a continuous field from a finite data set by approximating the posterior probability $\mathcal{P}(\varphi|d)$.  In fact, in both cases, Bayesian parameter estimation and Bayesian field inference, it is in general not necessary to compute the evidence, as long as one is interested in computing posterior expectation values. In contrast in model comparison, the evidence is the main measure of interest. In model comparison, multiple models, each with its parameters and assumptions, are compared by the probability of a model $\mathcal{M}_i$ given the data,
\begin{align}
\mathcal{P}(\mathcal{M}_i|d) = \frac{\mathcal{P}(d|\mathcal{M}_i)\mathcal{P}(\mathcal{M}_i)}{\mathcal{P}(d)}, \label{eq:modelcomp}
\end{align}
with $\mathcal{P}(d)= \sum_j \mathcal{P}(d|\mathcal{M}_j) \mathcal{P}(\mathcal{M}_j)$. Assuming that all models have the same a priori probability, this leads to a comparison of the evidences $\mathcal{P}(d|\mathcal{M}_j)$ between a set of different models $\{\mathcal{M}_j\}$. The ratio of the evidences of two models is called the Bayes factor, giving the betting odds for or against one of the models compared to the other. \\
A number of algorithms for Bayesian parameter estimation already exits, which can be divided into two classes of posterior estimation approaches. One approach approximates the posterior, i.e. it tries to find an analytic distribution that is close to the true posterior. The main approach here is \ac{VI}, which minimises a distance measure between the analytic distribution and the posterior distribution, for example the \ac{KL} (\citep{10.1214/aoms/1177729694}), and is often used in field inference. The other approach aims to generate a set of samples of the posterior distribution. This set of samples can be used to approximate the true posterior. The most popular posterior sampling algorithm is \ac{MCMC}. A summary of the basics of \ac{MCMC} and different implementations is given in \citep{Hogg_2018}. MCMC methods draw samples directly from the posterior, given a likelihood and a prior model. The simplest algorithm for \ac{MCMC} is the Metropolis-Hastings algorithm \citep{2015arXiv150401896R}, which gives a biased random walk through the parameter space depending on a proposal function and the initialisation of the algorithm. Still, there are several challenges, such as the tuning of the proposal function and the initialisation, which lead to advanced \ac{MCMC} methods such as Ensemble Sampling \citep{lu2023ensemble}, Gibbs Sampling \citep{4767596} and Hamiltonian Monte Carlo \citep{betancourt2018conceptual}. \\
For model comparison (eq.~\eqref{eq:modelcomp}), instead we are interested in the integration of the likelihood over the prior, or in other words in the evidence calculation. Computing Bayesian evidences is challenging in many applications, as discussed in  \citep{buchner_nested_2021}. The biggest challenges are, high-dimensional posteriors with multiple, well-separated modes or plateaus and a high information gain from the prior to the posterior, increasing the amount of time the algorithm spends in the low posterior mass regime. A concise overview of the different approaches for integration is given in \citep{preuss_comparison_2007}. Two integration methods that smoothly contract the parameter space from the prior to the posterior are simulated annealing \citep{inproceedings} and \ac{NS} \citep{skilling_nested_2004, skilling_nested_2006}. While simulated annealing uses fractional powers of the likelihood to get from the prior to the posterior, \ac{NS} instead takes samples from slices of the posterior and recombines them at the end. In particular, \ac{NS} transforms the multidimensional problem of integrating the likelihood over the prior into a series of nested volumes defined by likelihood contours and the enclosed prior. In doing so, \ac{NS} is able to estimate the log-evidence and the posterior samples simultaneously. \\
As summarised in \citep{buchner_statistical_2023} there are
several challenges in \ac{NS}. First, the computational cost of \ac{NS} depends on the choice of prior, i.e. the broader the prior, the higher the computational cost. Second, sampling from the likelihood restricted prior is not trivial. 
Finally, the rate at which the posterior is integrated is a stochastic quantity for which there is only a probabilistic description. 
Accordingly, many improvements to \ac{NS} have been proposed to address some of these challenges. A number of studies have focused on improving the calculation of evidence for likelihoods with peculiar shapes, such as likelihood plateaus. Likelihood plateaus violate the assumption of uniformity in sampling and lead to ambiguity in the ranking of samples, which often leads to an underestimation of the prior volume contraction and thus to an overestimation of the evidence. Accordingly, \citep{schittenhelm_nested_2021} propose a preprocessing step to correctly handle the plateaus in the likelihood. The main idea is to decompose the parameter sets into disjoint subsets that divide into plateaus and parts on which the usual \ac{NS} can be performed. Another approach to dealing with plateaus during \ac{NS}, without preprocessing, is given by \citep{Fowlie_2021}.
Other studies have aimed to work with a variable rate at which the posterior is integrated, known as dynamic \ac{NS} \citep{Higson_2018, speagle_dynesty_2020}. Dynamic \ac{NS} is particularly useful for parameter estimation, as standard \ac{NS} spends a lot of computational effort navigating to the posterior peak. Ultimately, dynamic \ac{NS} allows more samples to be placed in regions where we want higher resolution, and less in less interesting regions. 
However, most of the progress has been focused on improving the sampling process for the \ac{LRPS}. There are generally two different approaches that focus on \ac{LRPS} - rejection sampling, used by MultiNest \citep{Feroz_2009}, and chain-based sampling using Markov chains, implemented for example in PolyChord \citep{Handley_2015}. In addition, \citep{salomone_unbiased_2023} noted that standard \ac{NS} assumes independent prior samples given the likelihood constraint. However, this is usually not the case, leading to a bias in the evidence calculation. Accordingly, they introduce \ac{NS-SMC} based on the idea of importance sampling, which does not require the imprecise assumption of independent samples. \\
The dominant error in the evidence calculation, which is based on the statistical estimate of the shrinkage ratio $t_i$, can be reduced by taking a larger number of samples, which is where improvements in \ac{LRPS} focus. We take an orthogonal approach and try to increase the accuracy of \ac{NS} through a post-processing step that reduces the statistical error in each of the compression factors. To do this, we use \ac{IFT} to perform Bayesian field inference to reconstruct a continuous and smooth likelihood-prior-volume function given the likelihood contour information from \ac{NS}. The presented approach has been addressed in \citep{westerkamp_inferring_2023}. In this paper we aim to give a deeper introduction into the post-processing and perform further validation. \\
In the following, we first give a general overview of the methods used in section \ref{sec:MaterialsAndMethods}. This includes an introduction to \ac{NS} and its notation, an introduction to \ac{IFT} and the general explanation of one-dimensional correlated field inference in section \ref{sec:IFT}, and finally the method for inferring prior volume estimates and a possible implementation of it in section \ref{sec:BayesianInference}. In section \ref{sec:Results} we show the inference results given the implementation in section \ref{sec:BayesianInference} for two validation examples. In particular, we choose a Gaussian likelihood and a spike-and-slab likelihood \citep{mitchell_bayesian_2024}. Finally, we discuss the results, including an analysis of the computational cost, and conclude in section \ref{sec:Discussion}.

\section{Methods}
\label{sec:MaterialsAndMethods}
\subsection{Nested Sampling Algorithm}
\label{sec:standardnestedsampling}
\begin{figure}[!h]
  \centering
  \begin{subfigure}{0.49\textwidth}
  \includegraphics[width=\linewidth]{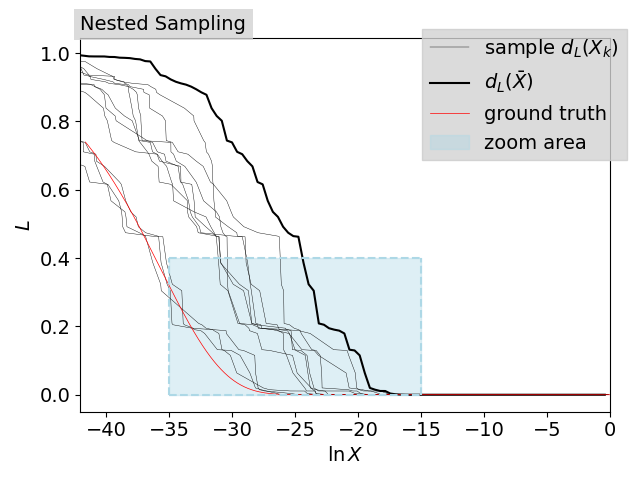}
  \caption{}
    \label{fig:GaussexampleFull}
    \end{subfigure}
     \begin{subfigure}{0.49\textwidth}
  \includegraphics[width=\linewidth]{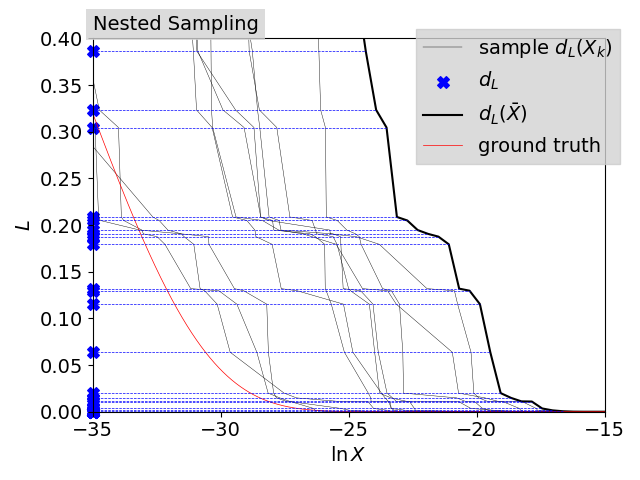}
  \caption{}
    \label{fig:GaussexampleZoom}
   \end{subfigure}
   \caption{Illustration of \acf{NS} output for simple Gaussian example introduced in \citep{skilling_nested_2006} and further elaborated in section \ref{sec:Gaussian} with two live points. The left side shows the full \ac{NS} data generated with \texttt{anesthetic} and the right side shows a zoomed in section, which is indicated on the left. The zoomed in image additionally shows the information of the likelihood dead contours $\vec{d}_L$, which we use as data for the Bayesian inference of the prior volumes. In both figures, the samples of likelihood-prior-volume functions defined by prior volume samples, $X_k$, $\vec{d}_L(X_k)$, are shown as well as a likelihood-prior-volume function defined by the deterministic \ac{NS} approach in eq.~\eqref{eq:MeanPriorVolume}, $\vec{d}_L(\bar{X})$.}
   \label{fig:GaussIllustrationExample}
\end{figure}

Using the notation introduced by \citep{skilling_nested_2004} the \ac{NS} likelihood is denoted via $\mathcal{L}(\theta):=\mathcal{P}(d|\theta)$ and the prior by $\pi(\theta):= \mathcal{P}(\theta)$. The evidence, $\mathcal{Z}:= \mathcal{P}(d)$ is calculated accordingly,  
\begin{align}
\mathcal{Z} = \int d\theta \mathcal{L}(\theta) \pi(\theta) \label{eq:evidence}.
\end{align}
The idea of \ac{NS} is to transform this possibly high-dimensional integral in parameter space into a one-dimensional one. Given the prior mass, $X$, enclosed by some likelihood contour $\mathcal{L}(\theta) = L$, 
\begin{align}
X(L) = \int_{\mathcal{L}(\theta)>L} d\theta ~\pi(\theta), \label{eq:priorvolume}
\end{align}
we can rewrite eq.~\eqref{eq:evidence} to a one-dimensional integral,
\begin{align}
\mathcal{Z} = \int_0^1 dX ~L(X)\label{eq:onedEvidence}, 
\end{align}
where $L(X)$ is the likelihood value on the $\theta$-contour that encloses the prior mass $X$ (eq. ~\eqref{eq:priorvolume}). 
The underlying algorithm for the calculation of the integral in eq. \eqref{eq:onedEvidence} can be summarised as follows: First, $n_\text{live}$ samples are drawn from the prior $\pi(\theta)$, which we call the live points.
For each of these samples the likelihood can be calculated. The sample with the lowest likelihood
is added to a new set, called the dead points, $\vec{d}_L := \{d_{L,i}\}$. A new sample
is drawn, restricted to the space of higher likelihood values (\ac{LRPS}).
Accordingly, we transfer samples from a set of live points to dead points with increasing likelihood, while adding new samples to the set of live points for which the likelihood values exceed the highest dead contour. This leads to the condition $d_{L, i} > d_{L,{i-1}}$. The prior volume under consideration shrinks at each iteration, $X_i < X_{i-1}$, by a compression factor $t_i$, 
\begin{align}
X_{i+1} = t_i X_i \label{eq:PriorVolumeCompression}.
\end{align}
Under the assumption that the samples are drawn from the prior independently within the highest dead contour, all compression factors, $\vec{t} = \{t_i\}$, are independent of each other and beta-distributed, $\mathcal{P}(t_i) = \text{Beta}(t_i|1,n_\text{live})$. The algorithm stops after $n_\text{iter}$ iterations. Finally, the set of dead-points and estimated prior volumes, defined by eq.~\eqref{eq:PriorVolumeCompression}, are used to approximate the evidence using the quadrature rule with the according weights $\omega_i$,
\begin{align}
\mathcal{Z} \approx Z = \sum_{i=1}^{n_\text{iter}} \omega_i d_{L,i}. \label{eq:ApproximatedEvidence}
\end{align}
In this study we use weights defined by the trapezoidal rule $\omega_i = \frac{1}{2}(X_{i-1} - X_{i+1})$ with $X_0 =1$ and $X_{n_\text{iter}+1} =0$.\\
In view of this procedure, \ac{NS} introduces a statistical uncertainty, since the prior volumes at each iteration, $X_i$, are not known, but only the distribution of the contraction factors defining the prior volumes is known. In the literature there are two different approaches for the estimation of the prior volume mentioned. The first one, we call it the statistical approach, samples $K$ chains of compression factors $\{t_i\}_k$ independently. Correspondingly, we can define several sets of prior volumes $\{X_i\}_k$, where for each chain $k = 1 , ..., N$ the prior volume at iteration $i$ is defined by the corresponding sets of contraction factors,$X_{i, k} = \prod_{j=1}^i t_{j,k}$.
The result is $K$ samples for the log-evidence using eq.~\eqref{eq:ApproximatedEvidence}, which allows us to obtain the mean estimate of the log-evidence and its uncertainty. The second approach, which we call the deterministic approach, instead gives no uncertainty estimate. Here, the mean of the log-compression factors $\langle \ln t_i \rangle_{\mathcal{P}(t_i)}= - 1/n_{\text{live}, i}$ is taken as an estimate, as discussed in \citep{feroz_importance_2019}.
This yields the deterministic prior volume estimation,
\begin{align}
\bar{X}_i = \prod_{j=1}^i \langle t_j \rangle_{\mathcal{P}(t_j)}= e^{\ln(\prod_{j=1}^i \langle t_j \rangle_{\mathcal{P}(t_j)})} \approx e^{\sum_{j=1}^i -1/n_{\text{live},j}} \label{eq:MeanPriorVolume},
\end{align}
and hence $\bar{X}_i \approx e^{-i/n_\text{live}}$ if the number of live points remains constant at each iteration (\citep{Handley_2015, Fowlie_2021}). In other words, the prior volume gets compressed exponentially. In figure \ref{fig:GaussIllustrationExample} we show the likelihood-prior-volume curves generated by \ac{NS} for a simple Gaussian example, which was introduced by \citep{skilling_nested_2006}. In section \ref{sec:Gaussian}, the details of this simple Gaussian case are discussed further. The according \ac{NS} likelihood contours were generated using the software package \texttt{anesthetic} \citep{anesthetic}. Figure \ref{fig:GaussexampleFull} shows the entire likelihood-prior-volume function generated together with the analytical ground truth. Figure \ref{fig:GaussexampleZoom} shows an enlarged section that is marked in the left panel. In \ac{NS} each likelihood dead contour, $d_{L,i}$, is accompanied by the estimated prior volume, described by all contraction factors up to the considered iteration. This leads to the corresponding \ac{NS} likelihood-prior-volume function, defined through either the statistical prior volume estimation, $\vec{d}_L(X_k)$ for $k=1,...,N$, or the deterministic prior volume estimation, $\vec{d}_L(\bar{X})$. Both panels of figure \ref{fig:GaussIllustrationExample} show both the statistical and the deterministic likelihood-prior-volume function. The zoomed in panel also shows the information on the likelihood contours given by \ac{NS}, which we will use as the only data, $\vec{d}_L$ for the inference of the likelihood-prior-volume function.

Despite \ac{NS} being mainly designed to estimate the evidence, it generates posterior samples $p_i$ at each iteration $i=1,...,n_\text{iter}$ of the model parameters $\theta_{\mathcal{M}}$ as a by-product as described in \citep{Handley_2015}, 
\begin{align}
p_i(\vec{t}) = \frac{\omega_i(\vec{t})d_{L,i}}{Z} = \frac{\omega_i(\vec{t})d_{L,i}}{\sum_i \omega_i(\vec{t}) d_{L,i}}\label{eq:posteriorsamples}.
\end{align}
These posterior samples can be used to compute the posterior expectation values of any function of the model parameters $g(\theta_\mathcal{M})$,
\begin{align}
\langle g(\theta_\mathcal{M}) \rangle_{\mathcal{P}(\theta_\mathcal{M}|d)} = \int d \theta_\mathcal{M} g(\theta_\mathcal{M}) \mathcal{P}(\theta_\mathcal{M}|d) \approx \frac{1}{n_\text{iter}} \sum_{i=1}^{n_\text{iter}} g(p_{i}). \label{eq:Sampleaveragedexpectation}
\end{align}
One such function and specific quantity of interest is the information gain from the prior to the posterior given by the \ac{KL},  
\begin{align}
H := D_\text{KL}(\mathcal{P}(\theta|d)|\pi(\theta)) = \int d \theta \mathcal{P}(\theta|d) \ln\biggl( \frac{\mathcal{P}(\theta|d)}{\pi(\theta)}\biggr) =  \int_0^1 dX \frac{L(X)}{Z} \ln \biggl( \frac{L(X)}{Z}\biggr). \label{KL}
\end{align}
The information gain is an important quantity for estimating the number of steps $n_\text{iter}$ needed to reach the posterior mass. It is therefore a good measure for determining a termination criterion. In particular, as noted in \citep{skilling_nested_2004, 10.1063/1.3703630}, the posterior set is reached after about $n_\text{live} H $ steps and the posterior is passed in $n_\text{live} \sqrt{C}$ steps, where $C$ is the number of dimensions. That is, more live points $n_\text{live}$ provide better sampling of the posterior, but also increase the time to reach the posterior, increasing the overall computation. In addition to the number of live points, $n_\text{live}$, the information gain $H$ and the average computational cost for \ac{LRPS} at each iteration and the average computational cost for evaluating the likelihood have an impact on the total computational cost $\mathcal{T}$. In particular, $\mathcal{T}$ scales as $\mathcal{O}(n_\text{live})$ and $\mathcal{O}(H^2)$ as pointed out in \citep{Ashton_2022}.
Looking at the error $\epsilon$ for the evidence calculation using \ac{NS}, we find according to \citep{chopin_properties_2010} that it is composed of three components under the assumption that we integrate up to the $n_\text{iter}$th iteration using eq.\eqref{eq:ApproximatedEvidence},
\begin{align}
\epsilon &= \sum_{i=1}^{n_\text{iter}} \omega_i d_{L,i} - \int_0^1 L(X) dX \\
&= \underbrace{-\int_0^{X_{n_\text{iter}}} L(X)dX}_{\text{truncation error}} + \underbrace{\biggl( \sum_{i=1}^{n_\text{iter}} \omega_i L(X_i) -\int_{X_{n_\text{iter}}}^1 L(X) dX\biggr)}_\text{numerical integration error} + \underbrace{\sum_{i=1}^{n_\text{iter}} \omega_i (d_{L,i}- L(X_i))}_\text{stochastic error}.
\end{align} 
According to \citep{speagle_dynesty_2020}, the numerical integration error introduced by replacing the integral by the trapezoidal rule is of the order of 
$\mathcal{O}(1/n_{\text{live}}^2)$ and therefore negligible as the number of live points goes to infinity. The truncation error occurs when we stop at a given maximum iteration $n_\text{iter}$. It can be kept small by choosing the stopping criterion wisely. Several approaches have been described in the literature to determine the final iteration $n_\text{iter}$, ranging from simultaneously computing the information $H$ to determine the location of the posterior set \citep{skilling_nested_2006}, to stopping as soon as the \ac{LRPS} becomes inefficient \citep{article} or as soon as the expected evidence from the remaining live points compared to the current evidence estimate is less than a user-defined tolerance \citep{Handley_2015, speagle_dynesty_2020, feroz_importance_2019}. \\
The error we are most interested in and aim to minimise is the stochastic error introduced by the unknown prior volumes. This error is of the order of $\mathcal{O}(n_\text{live}^{-\frac{1}{2}})$ (\citep{chopin_properties_2010}) and dominates the evidence approximation error.  As a result, the stochastic error goes to zero as the number of live points goes to infinity, but at the same moment the computational cost goes to infinity. Thus, there is a trade-off between the accuracy of the evidence calculation and the computation time. In the following we will present a post-processing step for \ac{NS}, that aims to reduce either the error in the evidence calculation or the time complexity, depending on the measure of interest. 
\subsection{Information Field Theory}
\label{sec:IFT}
\begin{figure}[H]
    \centering
    \caption*{\textbf{(a)} Change in $a$:}
     \vspace{1em}
    \begin{minipage}[b]{0.49\textwidth}
        \centering
        \includegraphics[width=\textwidth]{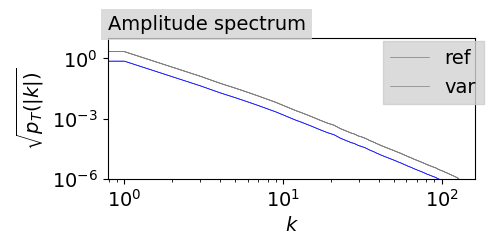}
    \end{minipage}
    \hfill
    \begin{minipage}[b]{0.49\textwidth}
        \centering
        \includegraphics[width=\textwidth]{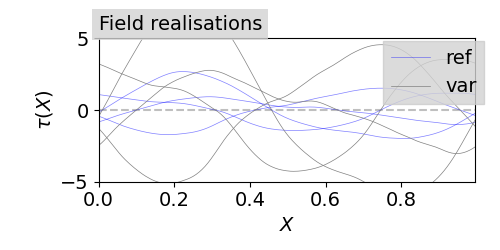}
    \end{minipage}

    \caption*{\textbf{(b)} Change in $m$:}
     \vspace{1em}

    \begin{minipage}[b]{0.49\textwidth}
        \centering
        \includegraphics[width=\textwidth]{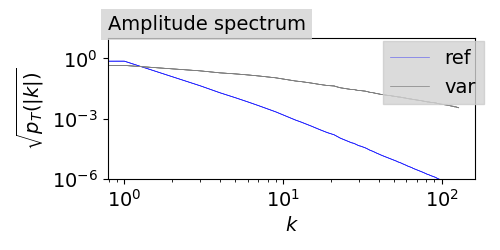}
    \end{minipage}
    \hfill
    \begin{minipage}[b]{0.49\textwidth}
        \centering
        \includegraphics[width=\textwidth]{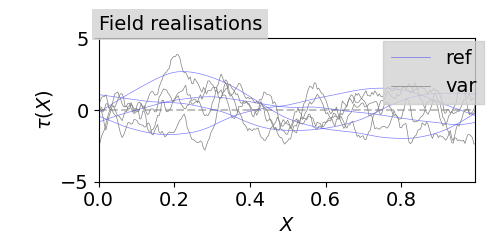}
    \end{minipage}
    
    \caption*{\textbf{(c)} Change in $\eta$:}
     \vspace{1em}

    \begin{minipage}[b]{0.49\textwidth}
        \centering
        \includegraphics[width=\textwidth]{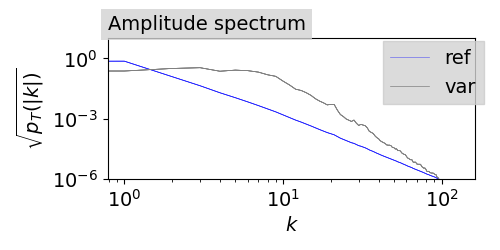}
    \end{minipage}
    \hfill
    \begin{minipage}[b]{0.49\textwidth}
        \centering
        \includegraphics[width=\textwidth]{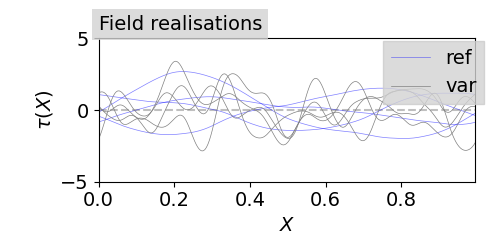}
    \end{minipage}
   \caption{Visualisation of the effect of changes in the mean value for one of the hyperparameters of the power spectrum model. The changed amplitude spectrum is shown on the left side and the according influence on the field realisation is shown on the right side by comparison of a reference correlated field (ref) and a variation of one of its hyper-parameters (var). Which hyper parameter is changed and how in comparison to the reference is denoted in table \ref{tab:CorrFieldHyperparameters}.}
   \label{fig:CorrelatedField}
\end{figure}

We use \ac{IFT} \citep{Ensslin_2019} for the joint reconstruction of the continuous likelihood-prior-volume function and the discrete set of prior volumes. \ac{IFT} focuses on  Bayesian field inference, or in other words, on the reconstruction of a continuous field from a discrete data set. Here we consider the likelihood-prior-volume function to be a one-dimensional field with an infinite number of degrees of freedom, which is to be reconstructed from a finite set of likelihood dead contours, $\vec{d}_L$. Thus, the inference problem is underconstrained, and we need prior knowledge of the likelihood-prior-volume function and the prior volumes to obtain the posterior. We call this prior probability distribution the joint reconstruction prior $\mathcal{P}(L(X), \vec{t})$ to avoid confusion with the prior volumes of \ac{NS}. The reconstruction likelihood $\mathcal{P}(\vec{d}_L| L(X), \vec{t})$, then is the probability of the measured likelihood dead contours given the likelihood-prior-volume function.
We join the information on the reconstruction prior and the reconstruction likelihood in Bayes theorem (eq.\eqref{eq:bayes}) to reconstruct the posterior probability of the field $L(X)$, which we call the reconstruction posterior $\mathcal{P}(L(X), \vec{t}| \vec{d}_L)$. This allows us to get any a  posteriori measure of interest, like for example the mean and the variance of the likelihood-prior-volume function. In Section \ref{sec:BayesianInference} we  will introduce the explicit reconstruction likelihood and prior models for the here introduced inference. Here, we focus on using \ac{IFT} and its software package \ac{NIFTy} (\citep{nifty5}) to implement a generative prior for the likelihood-prior-volume function, considering its correlation structure. \\
Since the prior model presented in section \ref{sec:BayesianInference} is based on Gaussian processes, this section describes Gaussian processes from the IFT perspective. Specifically we introduce a generative model for Gaussian processes with variable correlation structure. In other words, the aim is to generate a field, $\tau$, as a Gaussian process $\mathcal{G}(\tau, T)$ with an unknown covariance $T$. This implementation, which is desirable in many cases, ensures the smoothness of the likelihood-prior-volume function. We use this information in the presented algorithm to improve the accuracy of the evidence calculation. In the following sections, we will discuss the smoothness assumption and its positive and negative consequences in more detail. One approach for the implementation of a non-parametric model for fields with unknown correlation structure, was introduced in \citep{arras_variable_2022}. Analogously, we call this model the correlated field model, which suggests that not only the realisation of the field itself, $\tau$, is learned, but also the underlying correlation structure $T$. Here, we consider the simplest case of a one-dimensional correlated field. The correlated field is implemented as a generative process using the reparametrization trick introduced in \citep{kingma2015variational}, $\tau = A\xi_\tau$ with $T=AA^\dagger$ and $\mathcal{P}(\xi_\tau) = \mathcal{G}(\xi_\tau, \mathds{I})$. We can separate the field realisation from the field correlation structure using this basis transformation. This means, the new coordinates, $\xi_\tau$, have the same dimension as $\tau$, but are a priori uncorrelated. Assuming statistical homogeneity and isotropy, the covariance $T$ is fully defined by its power spectrum $p_T(|k|)$ via the Wiener-Khinchin theorem in Fourier space, 
\begin{align}
A_{kk^\prime} = (FAF^\dagger)_{kk^\prime} = 2\pi \delta(k-k^\prime) \sqrt{p_T(|k|)},
\end{align}
where $F$ is the Fourier transform and $\sqrt{p_T(|k|)}$ is the amplitude spectrum. The aim is to infer the power spectrum non-parametrically. This is achieved by building a model of the power spectrum where each hyperparameter is described by a Gaussian or log-normal prior with a given mean and standard deviation. Each hyperparameter, and thus the entire power spectrum, is learned during inference. More specifically,
the amplitude spectrum is implemented as an integrated Wiener process, a general continuous process, on the logarithmic scale $l = \log(|k|)$ for $k \neq 0$,
\begin{align}
\sqrt{p_T(l)} \propto e^{\gamma(l)}, \frac{d^2 \gamma}{dl^2} = \eta \xi_W(l), \mathcal{P}(\xi_W) = \mathcal{G}(\xi_W, \mathds{I}).
\end{align}
The integration gives 
\begin{align}
\gamma(l) = m l + \eta \int_{l_0}^l \int_{l_0}^{l^\prime} \xi_W(l^{\prime \prime}) dl^\prime dl^{\prime \prime}.
\end{align}
Here, $l_0$ is the first mode greater than zero and $m$ defines the slope of the integrated Wiener process, i.e. it is the slope of the amplitude spectrum on a double logarithmic scale. The parameter $\eta$ is called flexibility because it controls the total variance of the integrated Wiener process. In addition to these parameters, $m$ and $\eta$, which essentially determine the shape of the power spectrum, the total offset of the correlated field defined by the zero mode and another hyperparameter, called the fluctuations, $a$, which specifies the total fluctuations of the non-zero modes, are introduced. Overall, this gives a generative model for a Gaussian random field with unknown covariance. \\
The effect of changing $a$, $m$ and $\eta$ is shown in figure \ref{fig:CorrelatedField}. It shows a reference power spectrum and corresponding sample field realisations, together with the power spectrum and field realisations for a changed mean of one of the hyperparameters. The variances of the hyperparameters are the same for all cases and are kept small in order to better show the effect of the hyperparameters on the correlated field. The specific means and variances of the hyperparameters are listed in table \ref{tab:CorrFieldHyperparameters}. For the reconstruction itself, described in section \ref{sec:BayesianInference}, we keep the prior wide, which means that we take higher values for the standard deviations of each parameter to allow for flexibility of the model.\\ 
\ac{IFT} performs Bayesian field inference to infer the posterior for a continuous field given some data $d_\tau$. The exact relationship between the correlated field, $\tau$, and the likelihood-prior-volume function is discussed in section \ref{sec:BayesianInference}. Thereby, the posterior probability $\mathcal{P}(\tau|d_\tau)$ is approximated by a simpler posterior distribution $\mathcal{Q}(\tau|d_\tau)$ using \ac{VI}. The approximation is done by minimising the cross entropy term of the \ac{KL} between the actual posterior and its approximation $D_\text{KL}(\mathcal{Q}(\tau|d_\tau)|\mathcal{P}(\tau|d_\tau))$. In particular, we use the \ac{geoVI} introduced by \citep{Frank_2021}. The \ac{geoVI} algorithm optimises the cross-entropy of the \ac{KL} with respect to a non-linear normalising coordinate transformation that maps the posterior onto a standard Gaussian. This allows it to approximate non-Gaussian posteriors.
All numerics related to \ac{IFT} are implemented in the corresponding software package NIFTy \citep{nifty5}.

\begin{table}[H] 

\caption{Hyperparameters for correlated field samples shown in the figure \ref{fig:CorrelatedField}. The reference parameters are denoted by an index \textbf{r}. The other indices correspond to the labels of the sub-figures (\textbf{a}, \textbf{b}, \textbf{c}). Modified hyperparameter means with respect to the reference field are marked in blue.}
\newcolumntype{C}{>{\centering\arraybackslash}X}
\begin{tabularx}{\textwidth}{C|CCC|CCC|CCC|CCC}
\toprule
 & $\eta_\textbf{r}$ & $m_\textbf{r}$ & $a_\textbf{r}$ & $\eta_\textbf{a}$ & $m_\textbf{a}$ & $a_\textbf{a}$ & $\eta_\textbf{b}$ & $m_\textbf{b}$ & $a_\textbf{b}$ & $\eta_\textbf{c}$ & $m_\textbf{c}$ & $a_\textbf{c}$\\
\midrule
Mean & $0.5$ & $-6$ & $1.0$ & $0.5$ & $-6$ & \textcolor{blue}{$3.0$} & $0.5$ & \textcolor{blue}{$-2$} & $1.0$ & \textcolor{blue}{$10.0$} & $-6$ & $1.0$\\
Std & $0.5$ & $1^{-16}$ & $0.5$ & $0.5$ & $1^{-16}$ & $0.5$ & $0.5$ & $1^{-16}$ & $0.5$ & $0.5$ & $1^{-16}$ & $0.5$\\
\bottomrule
\end{tabularx}
\label{tab:CorrFieldHyperparameters}
\end{table}

\subsection{Bayesian Inference of the likelihood-prior-volume Function}
\label{sec:BayesianInference}
\begin{figure}[!h]
  \centering
  \begin{subfigure}{0.48\textwidth}
    \caption{}
         \vspace{1em}
  \includegraphics[width=\linewidth]{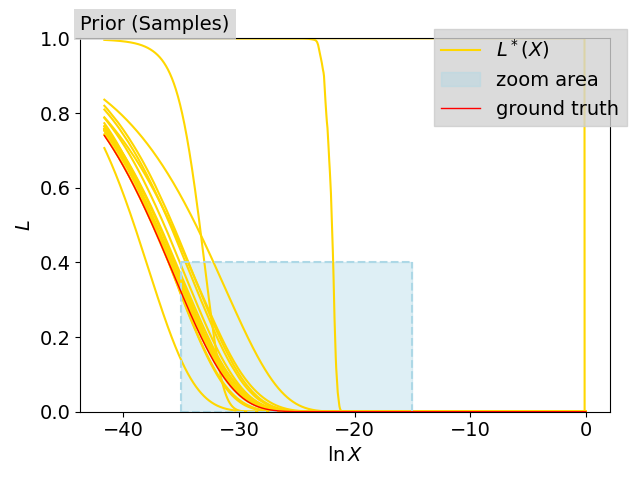}
    \label{fig:IFTPrior}
    \end{subfigure}
     \begin{subfigure}{0.48\textwidth}
       \caption{}
            \vspace{1em}
  \includegraphics[width=\linewidth]{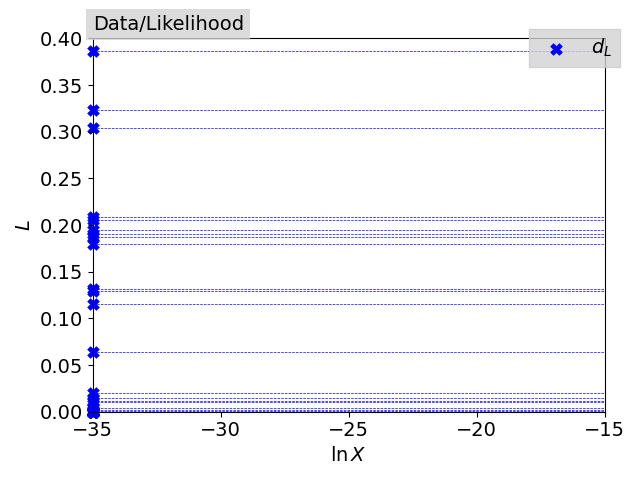}
    \label{fig:IFTLikelihood}
   \end{subfigure}\\
        \begin{subfigure}{0.48\textwidth}
          \caption{}
               \vspace{1em}
  \includegraphics[width=\linewidth]{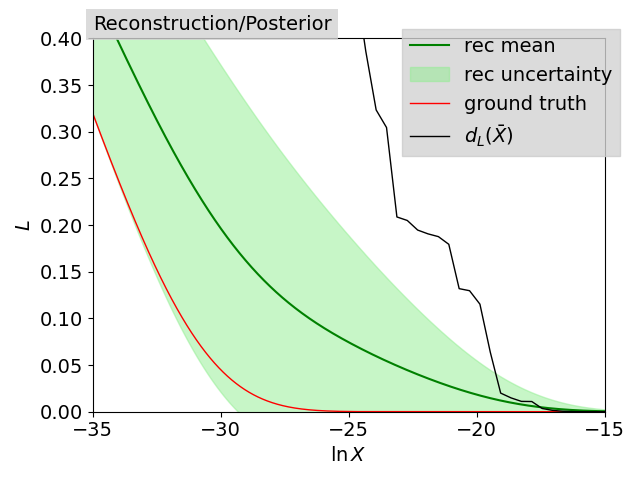}
    \label{fig:IFTPosteriorZoom}
   \end{subfigure}
        \begin{subfigure}{0.48\textwidth}
          \caption{}
               \vspace{1em}
  \includegraphics[width=\linewidth]{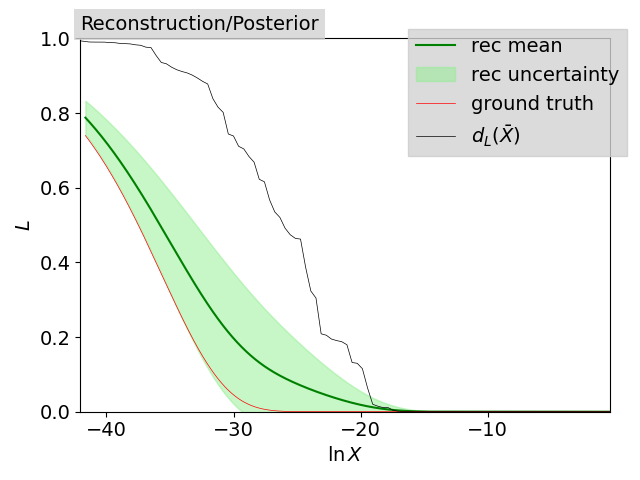}
    \label{fig:IFTPosterior}
   \end{subfigure}
   \caption{Illustration of the Bayesian field inference process for the simple Gaussian, which is further elaborated in section \ref{sec:Gaussian} for two live points. The prior samples, the data used and the final reconstruction compared to the \ac{NS} approach and the ground truth are shown. Figure \ref{fig:IFTPrior}: Prior samples for the likelihood-prior-volume function ($L^*(X)$) in yellow together with the ground truth. Besides a zoom area is marked (the same area as in figure \ref{fig:GaussexampleFull}) which is taken to zoom into the data in figure \ref{fig:IFTLikelihood} and the reconstruction in figure \ref{fig:IFTPosteriorZoom}. Figure \ref{fig:IFTLikelihood}: Data on likelihood dead contours for the given zoom area. Figure \ref{fig:IFTPosteriorZoom} and figure \ref{fig:IFTPosterior}: Reconstruction mean (rec mean) of the likelihood-prior-volume function and the associated uncertainty, defined via the onesigma contours (rec uncertainty), zoomed in \ref{fig:IFTPosteriorZoom} and full image in \ref{fig:IFTPosterior}. Moreover, the result for the likelihood-prior-volume function for the deterministic \ac{NS} approach is shown ($d_{L}(\bar{X})$), which  is the same as in figure \ref{fig:GaussexampleZoom}, and the analytic likelihood-prior-volume-function (ground truth).}  
   \label{fig:IFTRec}
\end{figure}

As noted in \cite{westerkamp_inferring_2023}, we use the smoothness assumption for the likelihood-prior-volume curve to improve the evidence calculation in \ac{NS}. The description of the algorithm is given below using the simple Gaussian example introduced by \citep{skilling_nested_2006} for illustration. The full information from \ac{NS} for this case is presented in Figure \ref{fig:GaussexampleFull}. As mentioned in section \ref{sec:standardnestedsampling}, \ac{NS} generates data on the likelihood dead contours $\vec{d}_L$. These data points are marked in the figure \ref{fig:GaussexampleZoom} on the likelihood axis. However, to compute the evidence we would need the likelihood-prior-volume function, including information on the prior volumes. In the following, we aim to give an algorithm that aims to improve the overall estimate on the prior volume and thereby reduces the uncertainty in the evidence. \\
We propose to jointly infer the likelihood-prior-volume function $L(X)$ and the prior volumes $\{X_i\}$ at each iteration $i=1,...,n_\text{iter}$ using Bayesian field inference as described in section \ref{sec:IFT}. As the data we only take into account the information we get from \ac{NS} about the probability of dead contours, $\vec{d}_L$.
The composition of the reconstruction is shown in figure \ref{fig:IFTPrior}. As ingredients for Bayes theorem, we use the joint reconstruction prior model for the contraction factors and the likelihood-prior-volume function, as well as the reconstruction likelihood model, which in our case is defined fully by the data. Figure \ref{fig:IFTPrior} shows prior samples for the likelihood-prior-volume function described by the correlated field to be learned. The data obtained by \ac{NS} for the likelihood dead contours, which then describes the reconstruction likelihood, is shown in \ref{fig:IFTLikelihood}. Merging these two models and approximating the reconstruction posterior with \ac{VI} yields posterior samples of the likelihood-prior-volume function and the set of prior volumes. Figure \ref{fig:IFTPosterior} shows the computed mean and uncertainty for the reconstructed posterior likelihood-prior-volume function, as well as the function obtained by pure \ac{NS} and the analytic ground truth. \\
Below, we describe a method to enforce the smoothness assumption on the likelihood-prior-volume curve using the correlated field model introduced in section \ref{sec:IFT}. In appendix \ref{app:exact approach}, we present an alternative approach that requires no $\delta$-function approximation; however, this method is currently only applicable for a MAP estimate and not for VI. The here presented approach, implements the smoothness by describing the derivative of the likelihood by the prior volume as a log-normal process, which can be achieved by using a correlated field as described in section \ref{sec:IFT}. By using a log-normal rather than a Gaussian process for the derivative of the logarithmic likelihood- prior-volume relation, we ensure that the likelihood-prior-volume function is monotonic. 
As a consequence of the correlated field model described above, it would be desirable that
\begin{align}
- e^{-\tau(\ln X)} = \frac{d \ln L }{d \ln X} \approx \text{const},
\end{align}
with $\tau$ drawn from a Gaussian process.
However, it can be seen that for the simplest case, a Gaussian likelihood model, this assumption is not fulfilled. Accordingly, we introduce a reparametrization $f$ that maps $\ln L$ such that we find a damped log-normal process for $\frac{d \ln L }{d \ln X}$,
\begin{align}
\frac{d f_{\ln L}}{d \ln X} := \frac{d f(\ln L)}{d \ln X} = \frac{1}{\ln L_\text{max} - \ln L } \frac{d \ln L}{d \ln X} = - e^{-\tau(\ln X)}. \label{eq:repGaussianProc}
\end{align}
This way, a constant $\tau$ perfectly captures the Gaussian case, and non-Gaussianity is absorbed in excitations of $\tau$ around this constant, as described in section \ref{sec:IFT}. The derivation of the reparametrization for the Gaussian case is given in appendix \ref{app:Reparametrization}. Appendix \ref{app:Maximal Likelihood Calculation} shows how to calculate the corresponding $L_\text{max}$ according to \citep{handley_quantifying_2019} if it is not known analytically. The joint reconstruction prior for reparametrized likelihood-prior-volume function $f_{\ln L}$ and the contraction factors $\vec{t}$ is fully defined via the joint prior $\mathcal{P}(\tau, \vec{t})$,
\begin{align}
\mathcal{P}(f_{\ln L}, \vec{t}) = \mathcal{P}(f_{\ln L}| \vec{t})\mathcal{P}(\vec{t}) = \mathcal{G}(\tau, T) \prod_{i=1}^{n_\text{iter}} \text{Beta}(t_i|1,n_{\text{live}, i}) = \mathcal{P}(\tau, \vec{t}). \label{eq:JointPrior}
\end{align} 
The reconstruction likelihood is described by the solution of the differential equation in eq.~\eqref{eq:repGaussianProc}, which can be written as a function of the set of contraction factors,
\begin{align}
f_{\ln L} (\vec{t}) = f_{\ln L}(\ln X_j = \sum_{i=1}^j t_i) = f_{\ln L}(0) - \int_0^{\sum_{i=1}^j t_i} e^{-\tau(z)} dz. \label{eq:DiffEqSolution}
\end{align}
This leads to a likelihood model which is a $\delta$ function, which we approximate by a Gaussian with a small chosen variance $\sigma_\delta$, 
\begin{align}
\mathcal{P}(f(\ln \vec{d}_L)| \tau, \vec{t}) = \delta(f(\ln \vec{d}_L) - f_{\ln L}(\vec{t}))
\approx \mathcal{G}(f(\ln \vec{d}_L) - f_{\ln L}(\vec{t}), \sigma_\delta).
\label{eq:RecLikelihood}
\end{align}
The result is the joint reconstruction posterior,
\begin{align}
\mathcal{P}(\tau, \vec{t} |f(\ln \vec{d}_L)) \propto \mathcal{P}(f(\ln \vec{d}_L)| \tau, \vec{t}) \mathcal{P}(\tau, \vec{t}) \label{eq:RecPosterior} .
\end{align}
The posterior is approximated using \ac{geoVI} as described in section \ref{sec:IFT}, slowly increasing the number of samples from iteration to iteration. We choose $\sigma_{\delta} = 0.1~\text{min}(\text{dist}(f(\ln \vec{d}_L)))$ to ensure that the likelihood does not allow for the exchange of two data points or even non-monotonicity. Finally, the prior of $\tau$ prefers a constant flat $\tau$ without excitation, corresponding to a linear relation between $f_{\ln L}$ and $\ln X$, corresponding to a Gaussian likelihood in \ac{NS}. However, the prior of $\vec{t}$ prefers certain distances between the prior volumes given by the beta distribution. The likelihood ensures that the the reconstructed function $f_{\ln L}$ evaluated at the reconstructed prior volumes $X$  matches the values of $f(\ln \vec{d}_L)$. This means that, for example, if there is a jump in $f(\ln \vec{d}_L)$, $\mathcal{P}(\tau)$ prefers a corresponding jump in $\ln X$ to ensure smoothness and to avoid deviations from the linear reparametrised likelihood-prior-volume relation, while $\mathcal{P}(\vec{t})$ on average prefers an increase in $\ln X$ defined by the beta distribution.
\section{Results and Analysis}
\label{sec:Results}
For validation, we consider two cases: A simple Gaussian case as described in \citep{skilling_nested_2006} and a spike and slab likelihood as introduced in \citep{mitchell_bayesian_2024}. The according data on the likelihood live and dead contours is generated using the \texttt{anesthetic} package by \citep{anesthetic}. These test cases are valuable for checking the consistency of the presented method, as they allow the analytical calculation of evidence.
\subsection{Gaussian Case}
\label{sec:Gaussian}
\begin{figure}
  \centering
  \begin{subfigure}{0.48\textwidth}
    \caption{$n_\text{live} = 2$:}
     \vspace{1em}
    \includegraphics[width=\linewidth]{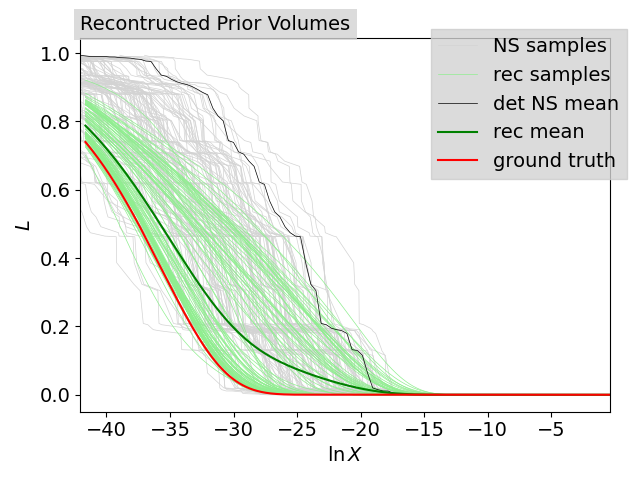}
  \end{subfigure}
  \begin{subfigure}{0.48\textwidth}
       \vspace{1em}
    \includegraphics[width=\linewidth]{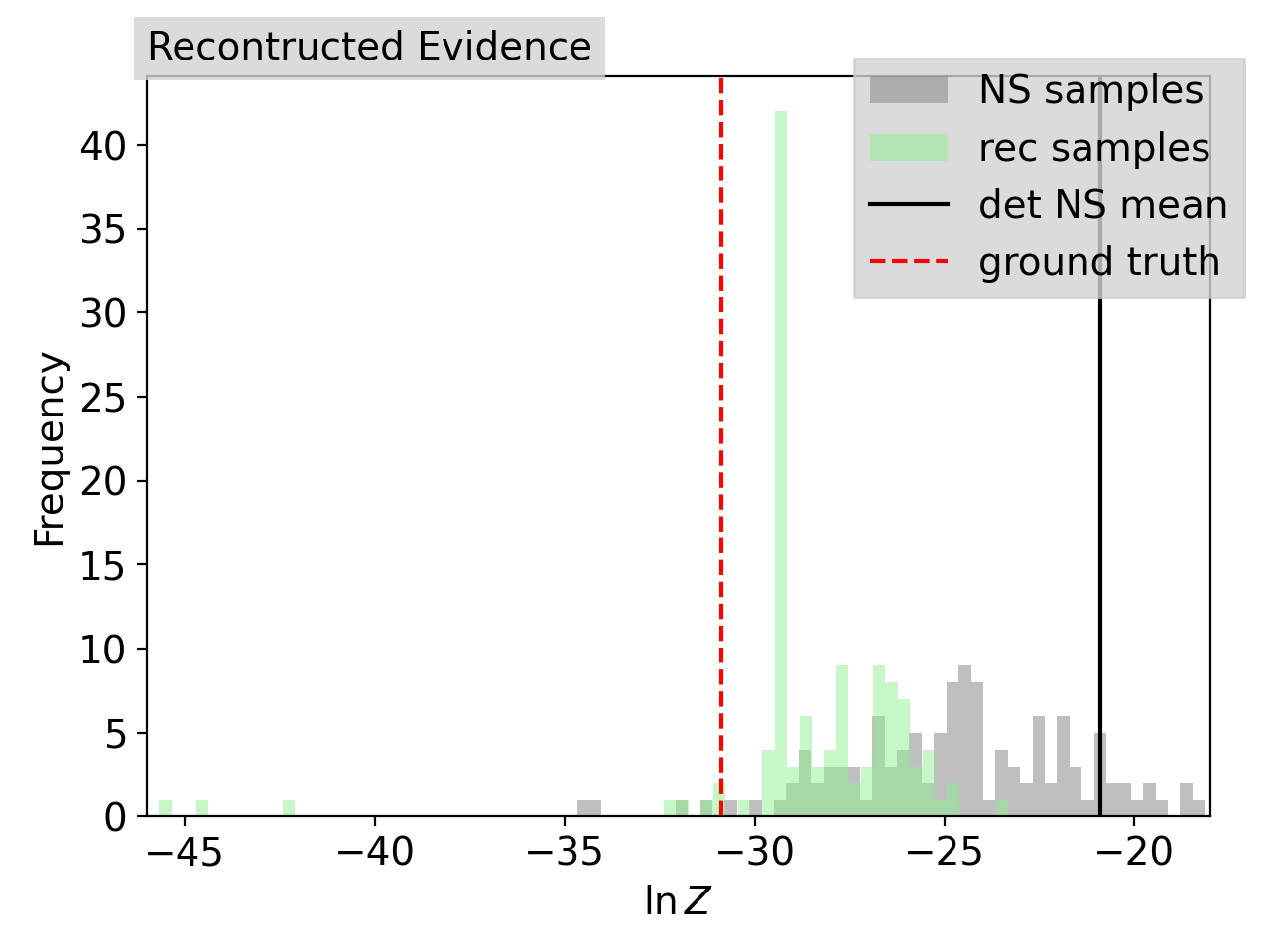}
  \end{subfigure}\\
    \begin{subfigure}{0.48\textwidth}
        \caption{$n_\text{live} = 10$:}
     \vspace{1em}
    \includegraphics[width=\linewidth]{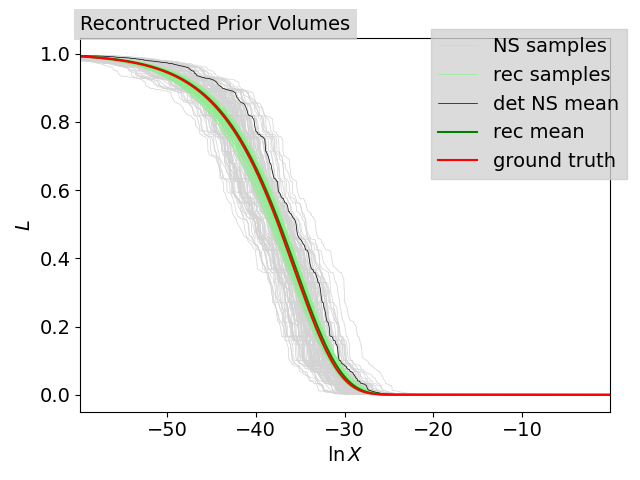}
  \end{subfigure}
  \begin{subfigure}{0.48\textwidth}
         \vspace{1em}
    \includegraphics[width=\linewidth]{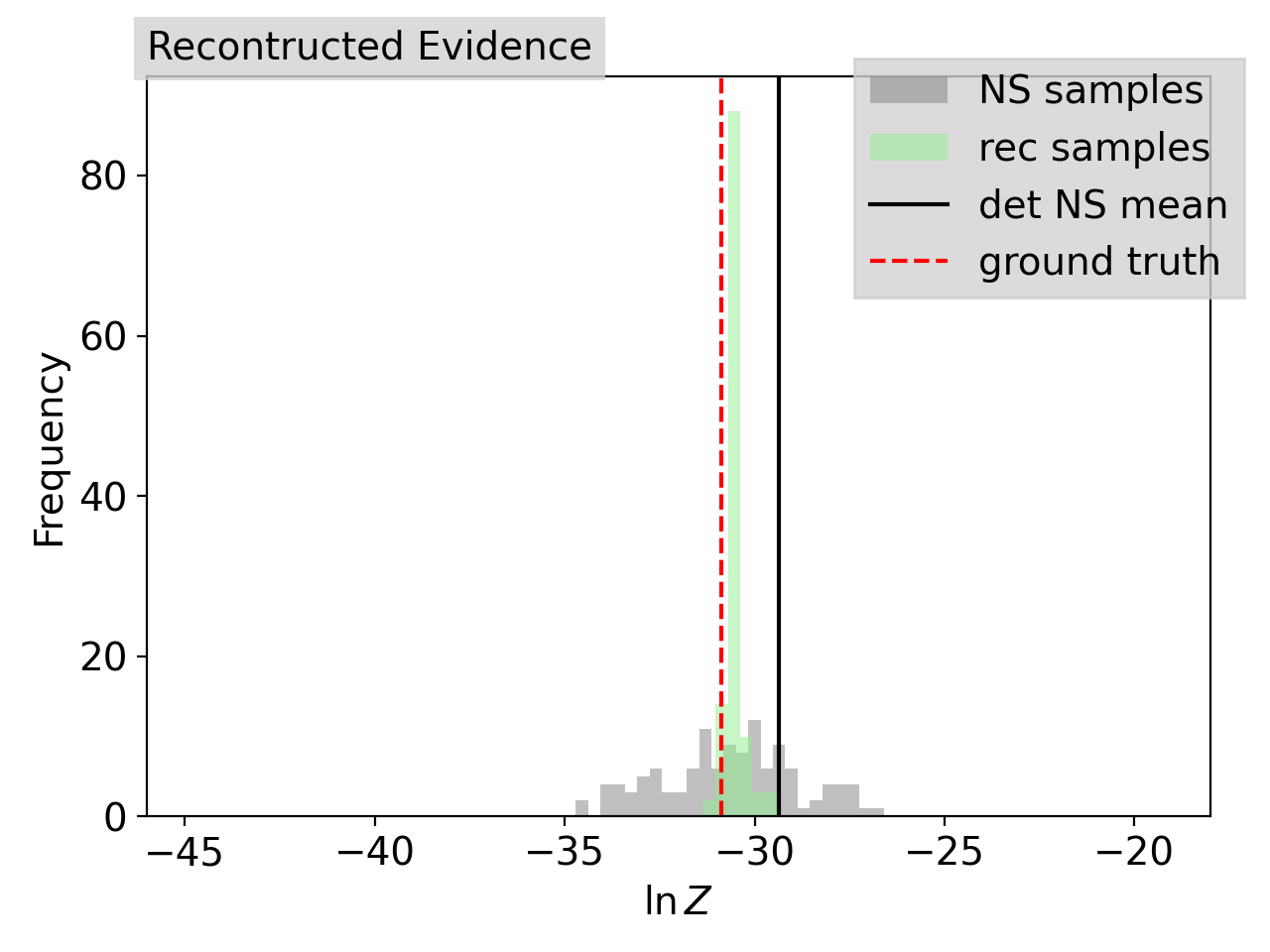}
  \end{subfigure}\\
    \begin{subfigure}{0.48\textwidth}
            \caption{$n_\text{live} = 1000$:}
                 \vspace{1em}
    \includegraphics[width=\linewidth]{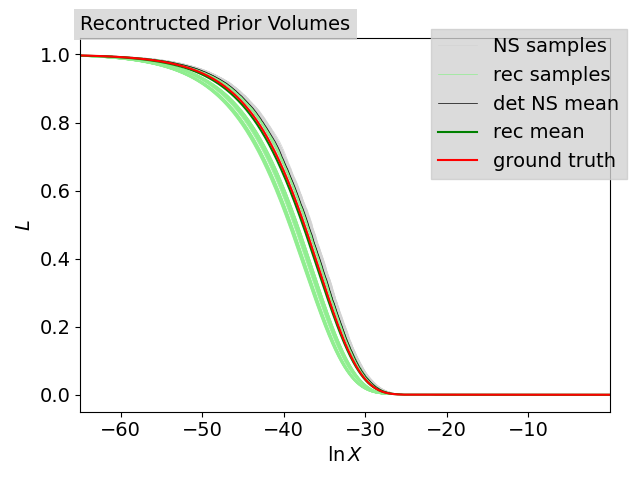}
  \end{subfigure}
  \begin{subfigure}{0.48\textwidth}
       \vspace{1em}
    \includegraphics[width=\linewidth]{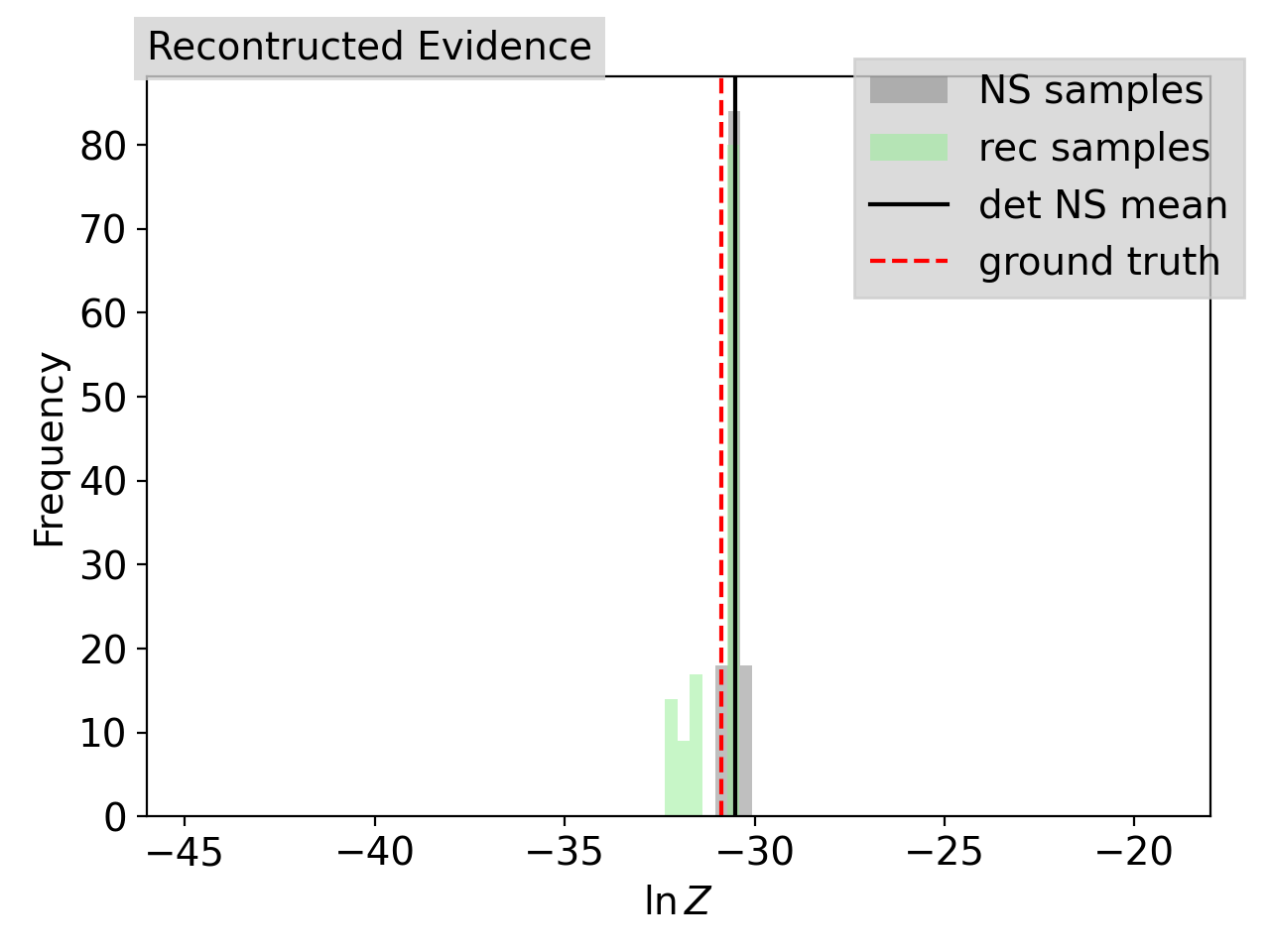}
  \end{subfigure}
  \caption{Reconstruction results for the Gaussian prior volumes accompanied by the likelihood contours given by \ac{NS} on the left and the computed log-evidence on the right for $n_\text{live} \in \{2, 10, 1000\}$ from top to bottom. The inferred posterior samples (rec samples) are shown together with their mean (rec mean) and compared with the corresponding statistical (\ac{NS} samples) and deterministic (det \ac{NS} mean) \ac{NS} results and the ground truth.}
  \label{fig:GaussRecLikelihoodPriorVolumeFunction}
\end{figure}

As a first validation test case, we use a zero-centred Gaussian likelihood,
\begin{align}
L(\theta) = \exp\biggl( -\frac{r^2}{2 \sigma^2}\biggr) ~~~ \text{with} ~~~ r^2=\sum_{i=1}^C \theta_i^2, ~~ \theta = \{\theta_i\}_{i=1,...,C}\label{eq:simpleGaussianLikelihood}
\end{align}
in $C=10$ dimensions with as variance $\sigma=0.02$. As a prior we use, in analogy to \citep{skilling_nested_2006}, a flat prior on the unit sphere
\begin{align}
\pi (\theta) = \frac{C/2!}{\pi^{C/2}} ~~~ \text{with} ~~~ r<1. \label{eq:simpleGaussianPrior}
\end{align}
The evidence given the probability and prior above can be calculated analytically to be
\begin{align}
\mathcal{Z} = \int_{-\infty}^\infty d\theta^C ~ L(\theta) \pi(\theta)  =\frac{(C/2)!}{(2\sigma^2)^{C/2}}. \label{eq:gauss_Z_gt}
\end{align}
The definition of the prior mass in $C$ dimensions is given by $X=r^C$, which allows us to compute the ground truth of the likelihood-prior-volume function, 
\begin{align}
L(X) = \exp\biggl(\frac{-X^{2/C}}{2\sigma^2}\biggr). \label{eq:gauss_L_gt}
\end{align}
Figure \ref{fig:GaussRecLikelihoodPriorVolumeFunction} shows the ground truth likelihood-prior-volume function together with samples of the likelihood-prior-volume function defined by the reconstructed prior volumes or the statistical approximated prior volumes from \ac{NS} for a constant number of live points, $n_\text{live} \in \{2, 10, 1000\}$. Also shown is the likelihood-prior-volume function for the deterministic \ac{NS} approach using eq.~\eqref{eq:priorvolume}. As described in section \ref{sec:standardnestedsampling}, it can be seen that the standard deviation for the \ac{NS} approach decreases as the number of live points increases. For each of these prior volume estimation approaches, the statistical \ac{NS}, the deterministic \ac{NS} or the \ac{IFT} based, we calculate the log-evidence using the weighted sum in eq.~\eqref{eq:ApproximatedEvidence}. This gives us sample sets of evidence for the statistical \ac{NS} and \ac{IFT} approaches, which are plotted as a histogram in figure \ref{fig:GaussRecLikelihoodPriorVolumeFunction} together with the analytical ground truth and the deterministic \ac{NS} approach. The corresponding results for each of the approaches for the mean and, where applicable, the standard deviation are given in table \ref{tab:GaussEvidence}. Further discussion of the results can be found in section \ref{sec:Discussion}.

\begin{table}[H] 
\caption{Inferred (\ac{IFT}) and \ac{NS} (NS stat: statistical, NS det: deterministic) results for the computed Gaussian log-evidence, represented by the mean and the according standard deviation for $n_\text{live} \in \{2, 10,1000\}$. The ground truth is $\ln Z = -30.87$. The histograms of the distributions of the evidences are shown in figure \ref{fig:GaussRecLikelihoodPriorVolumeFunction}.}
\newcolumntype{C}{>{\centering\arraybackslash}X}
\begin{tabularx}{\textwidth}{C|CCC|CC}
\toprule
$n_\text{live}$ & \multicolumn{3}{c}{mean}  &   \multicolumn{2}{c}{standard deviation} \\
 & IFT & NS stat & NS det &  IFT & NS stat\\
\midrule
$2$ & $-28.51$ & $-25.11$ & $-20.90$ &$2.98$  & $3.23$\\
$10$ & $-30.56$ & $-30.44$ & $-29.37$ & $0.24$ & $1.6$\\
$1000$ & $-31.01$ & $-30.52$ & $-30.51$ & $0.61$  & $0.17$\\
\bottomrule
\end{tabularx}
\label{tab:GaussEvidence}
\end{table}

\subsection{Spike \& Slab Case}
\label{sec:spikeandslab}
As a next step we consider a non-Gaussian test case for validation. In particular, we will look at a spike and slab likelihood known from Bayesian variable selection.
\citep{mitchell_bayesian_2024}, which is the sum of a zero-centred spike and a broad Gaussian background. This leads to an abrupt change in the prior volume $X$ with increasing likelihood. The corresponding likelihood is given by

\begin{align}
L(\theta) = a \exp\biggl(-\frac{r^2}{2\sigma_1^2}\biggr) + (1-a) \exp\biggl(-\frac{r^2}{2\sigma_2^2}\biggr) ~~~\text{with}~~~r^2=\sum_{i=1}^C \theta_i^2. 
\end{align}
Again, we choose a flat prior as described in eq. \eqref{eq:simpleGaussianPrior}. Accordingly, we are able to calculate the evidence analytically, which gives us a good point for comparison, 

\begin{align}
\mathcal{Z} = \int_{-\infty}^{\infty} d\theta^C L(\theta) \pi(\theta)= C/2! \biggl(a (2\sigma_1^2)^{C/2} + (1-a) (2\sigma_2^2)^{C/2} \biggr). \label{eq:spikeandslab_Z_gt}
\end{align}

Just as in section \ref{sec:Gaussian}, we can get the analytic likelihood-prior-volume-function given the prior volume $X=r^C$,
\begin{align}
L(X) = a \exp\biggl(\frac{-X^{2/C}}{2\sigma_1^2}\biggr) + (1-a) \exp\biggl(\frac{-X^{2/C}}{2\sigma_2^2}\biggr). \label{eq:SpikeAndSlabLikelihood}
\end{align}
The parameters of the spike and slab likelihood under consideration are denoted in Table \ref{tab:Spikaandslabparams}.

\begin{figure}
  \centering
  \begin{subfigure}{0.48\textwidth}
      \caption{$n_\text{live} = 2$:}
     \vspace{1em}
    \includegraphics[width=\linewidth]{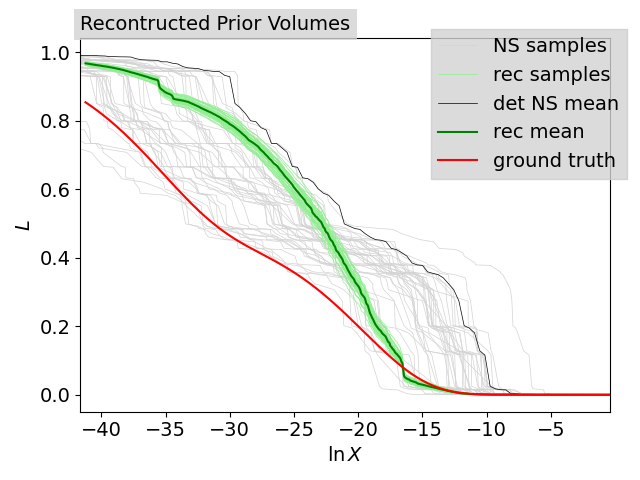}
  \end{subfigure}
  \begin{subfigure}{0.48\textwidth}
       \vspace{1em}
    \includegraphics[width=\linewidth]{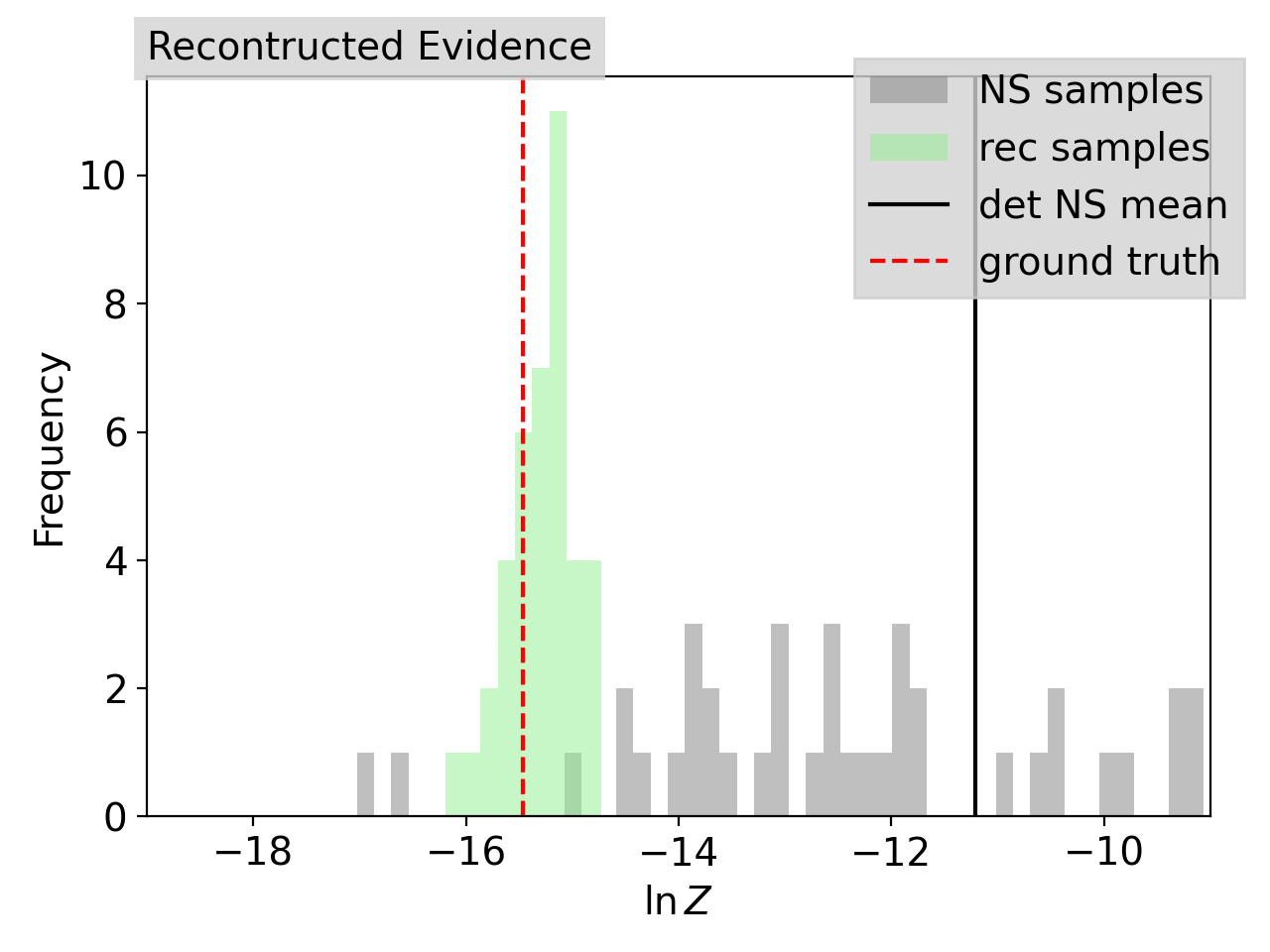}
  \end{subfigure} \\
    \begin{subfigure}{0.48\textwidth}
      \caption{$n_\text{live} = 10$:}
     \vspace{1em}
    \includegraphics[width=\linewidth]{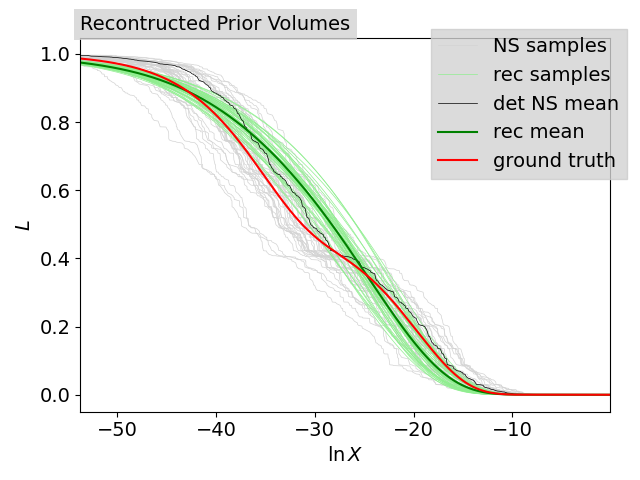}
  \end{subfigure}
  \begin{subfigure}{0.48\textwidth}
       \vspace{1em}
    \includegraphics[width=\linewidth]{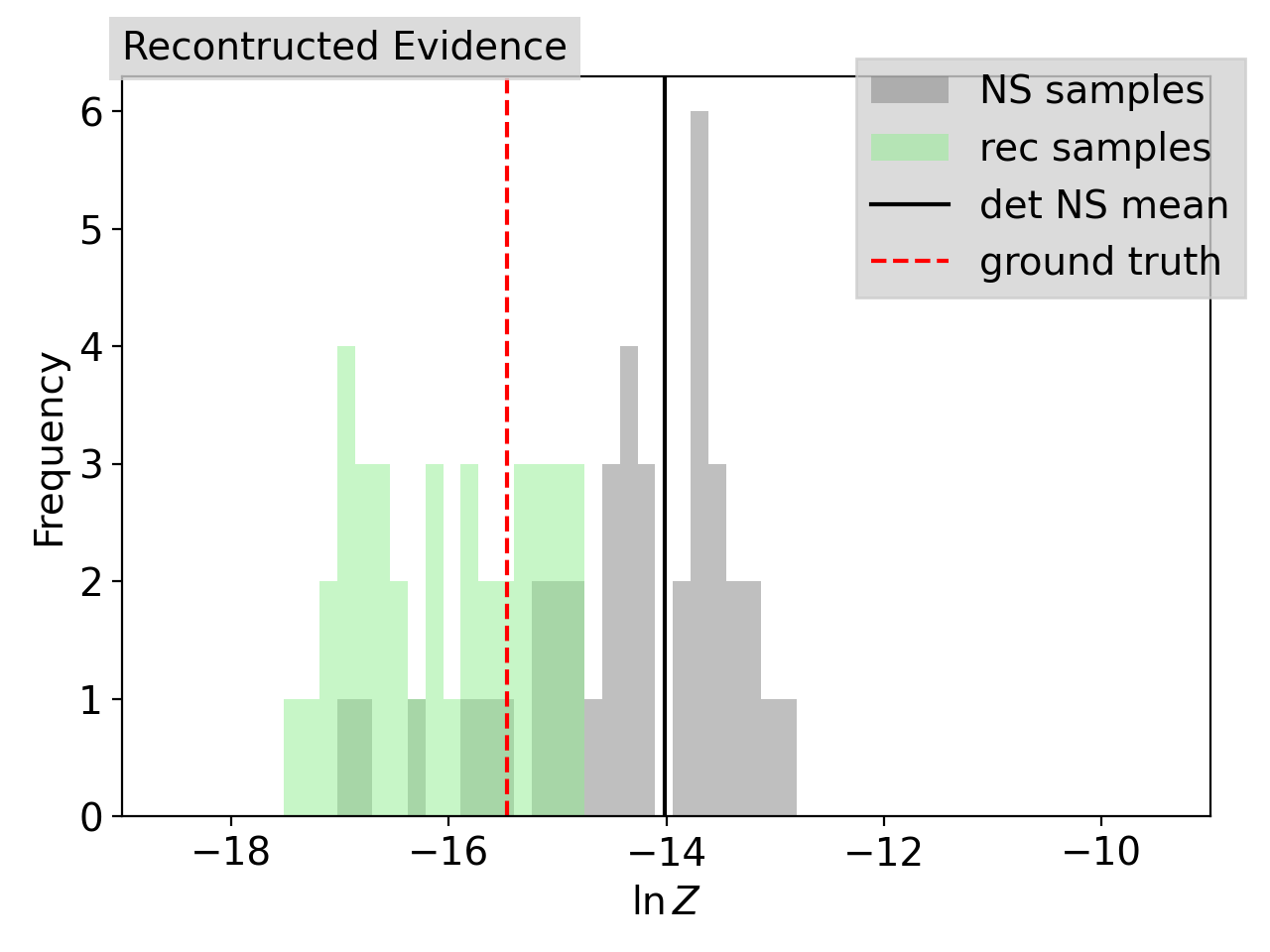}
  \end{subfigure}\\
      \begin{subfigure}{0.48\textwidth}
      \caption{$n_\text{live} = 1000$:}
     \vspace{1em}
    \includegraphics[width=\linewidth]{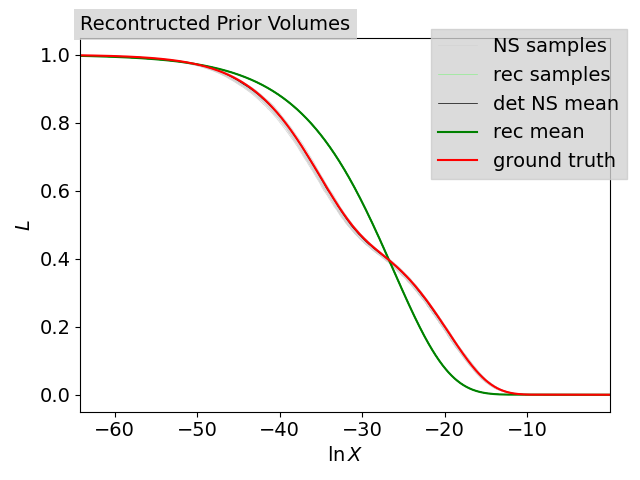}
  \end{subfigure}
  \begin{subfigure}{0.48\textwidth}
       \vspace{1em}
    \includegraphics[width=\linewidth]{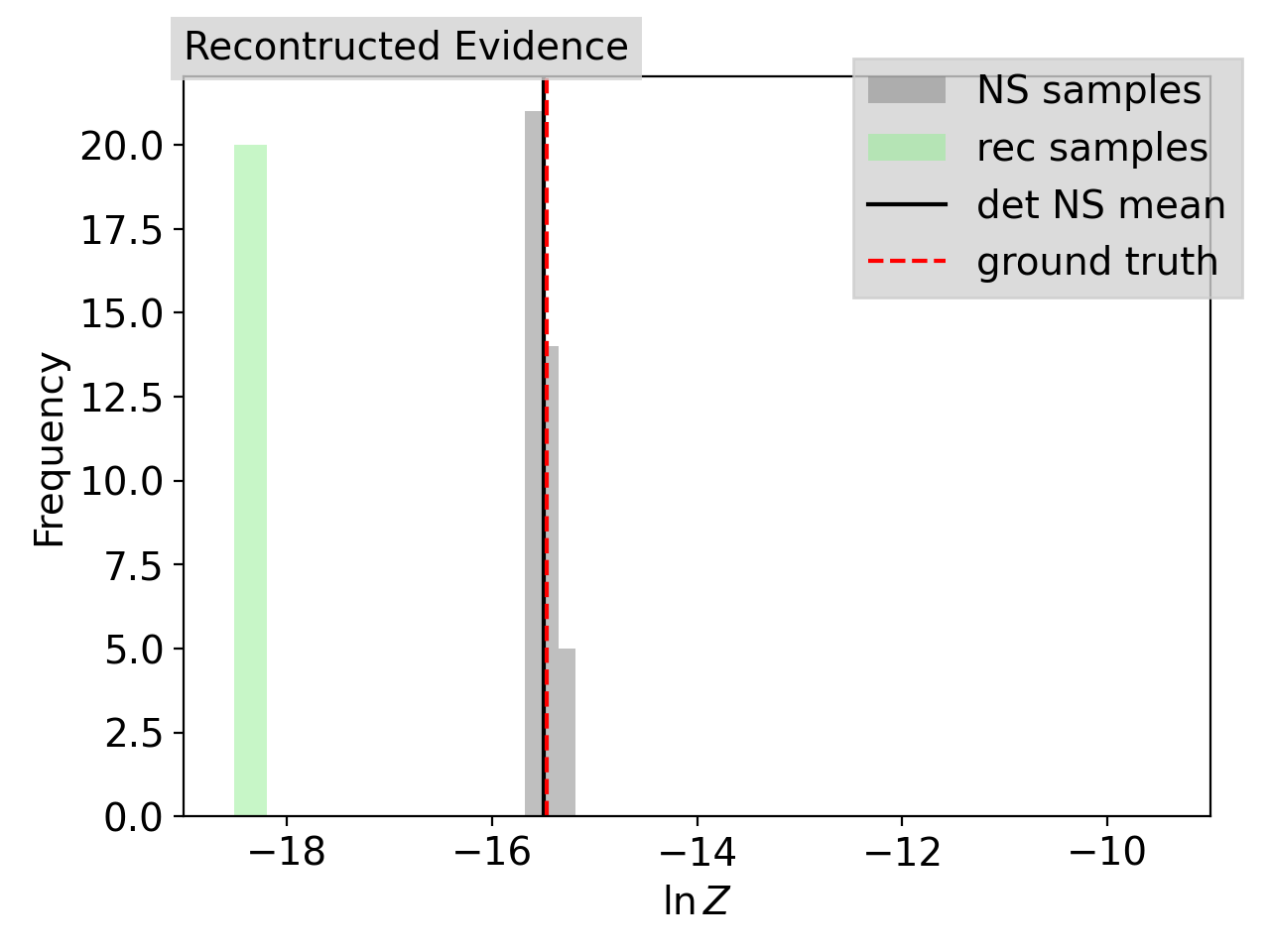}
  \end{subfigure}
  \caption{Reconstruction results for the spike-and-slab prior volumes accompanied by the likelihood contours given by \ac{NS} on the left and the computed log-evidence on the right for $n_\text{live} \in \{2, 10, 1000\}$ from top to bottom. The inferred posterior samples (rec samples) are shown together with their mean (rec mean) and compared with the corresponding statistical (\ac{NS} samples) and deterministic (det \ac{NS} mean) \ac{NS} results and the ground truth.}
    \label{fig:spike_and_slab_rec}
  \end{figure}

\begin{table}[H] 
\caption{Parameters for the spike and slab likelihood described in eq.~\eqref{eq:SpikeAndSlabLikelihood}.\label{tab:Spikaandslabparams}}
\newcolumntype{C}{>{\centering\arraybackslash}X}
\begin{tabularx}{\textwidth}{CCC}
\toprule
\textbf{Parameter}	& \textbf{Meaning} & \textbf{Value}\\
\midrule
$C$	& number of dimensions & 10\\
$a$ & relative weight of Gaussians & 0.5 \\
$\sigma_1$	& std of Gaussian weighted by $a$ & 0.1 \\
$\sigma_2$	& std of Gaussian weighted by $(1-a)$ & 0.02 \\
\bottomrule
\end{tabularx}
\end{table}
Figure \ref{fig:spike_and_slab_rec} shows the ground truth of the likelihood-prior-volume function (eq.~\eqref{eq:SpikeAndSlabLikelihood}) and evidence (eq.~\eqref{eq:spikeandslab_Z_gt}) together with the corresponding samples and mean (eq.~\eqref{eq:priorvolume}) given by the corresponding \ac{NS} runs for $n_\text{live} \in \{2, 10, 1000\}$. We infer the likelihood-prior-volume function and the set of prior volumes jointly using the inference algorithm described in section \ref{sec:BayesianInference}. As a result, we obtain a set of posterior samples on the likelihood-prior-volume functions and the prior volumes, which leads to a set of posterior samples of the evidence using eq.\eqref{eq:evidence}. The posterior samples for the likelihood contours as a function of the reconstructed prior volumes (rec samples) as well as their mean (rec mean) are shown on the left side of figure \ref{fig:spike_and_slab_rec}. The computed corresponding evidence is shown on the right side of figure \ref{fig:spike_and_slab_rec}. For comparison, additionally the results for the statistical \ac{NS} approach and the deterministic \ac{NS} approach as well as the ground truth are shown. The computed mean evidences for the sample sets for the \ac{IFT} and the \ac{NS} approach are listed in table \ref{tab:SpikeandSlabEvidence} together with the corresponding standard deviation. A further discussion of the results is given in section \ref{sec:Discussion}.
\begin{table}[H] 
\caption{Inferred (\ac{IFT}) and \ac{NS} (NS stat: statistical, NS det: deterministic) results for the computed spike-and-slab log-evidence, represented by the mean and the according standard deviation for $n_\text{live} \in \{2, 10,1000\}$. The ground truth is  $\ln Z = -15.47$. The histograms of the distributions of the evidences are shown in figure \ref{fig:spike_and_slab_rec}}
\newcolumntype{C}{>{\centering\arraybackslash}X}
\begin{tabularx}{\textwidth}{C|CCC|CC}
\toprule
$n_\text{live}$ & \multicolumn{3}{c}{mean}  &   \multicolumn{2}{c}{standard deviation} \\
 & IFT & NS stat & NS det &  IFT & NS stat \\
\midrule
$2$ & $-17.26$&  $-13.00$  & $-11.21$ & $1.67$ & $1.91$\\
$10$ & $-16.03 $ & $-14.62$ & $-14.01$& $0.78$ & $0.80$\\
$1000$ & $-18.35$ & $-15.49$ & $-15.50$ &$0.01$ & $0.10$ \\
\bottomrule
\end{tabularx}
\label{tab:SpikeandSlabEvidence}
\end{table}

\section{Discussion}
\label{sec:Discussion}
In \ac{NS}, the statistical error in one of the contraction factors $t_i$ affects each upcoming prior volume according to eq.~\eqref{eq:PriorVolumeCompression}. For this reason it has a major impact on the calculation of the logarithmic evidence. However, the propagation of error does not occur on the likelihood contour information, which is assumed to be accurate, but only on the prior volume estimates. We use additional knowledge to be able to assign an improved estimate of the prior volume to the corresponding likelihood contour. The assumption we take into account is the smoothness of the likelihood-prior-volume function, which is valid for a large set of problems. Of course, if a problem is considered where the likelihood-prior-volume function is not smooth, this post-processing algorithm will not be applicable and the inference of the prior volumes could even make the result worse. One extreme example, that was discussed in \citep{Fowlie_2021}, is the the wedding cake likelihood. Furthermore, the inference algorithm has its limits when dealing with likelihood plateaus. The reason for this limitation is that we model the derivative of the likelihood-prior-volume function as a modified log-normal process. A zero slope would therefore correspond to an infinite excitation of the correlated field $\tau$. A way to deal with this problem was suggested by \citep{schittenhelm_nested_2021}. Accordingly one could split the data set into several parts and perform the inference solely on the strictly negative monotonic regions of the likelihood-prior-volume functions. \\
In terms of computational cost, our overall goal is to define a post-processing algorithm whose computational cost is independent of the number of live points, $n_\text{live}$. As described in section \ref{sec:standardnestedsampling}, the computational cost of a \ac{NS} run is proportional to the number of live points $n_\text{live}$ and the \ac{KL}, $H$, while the error in $\ln Z$ is proportional to $\sqrt{H}$ and anti-proportional to $\sqrt{n_\text{live}}$. Therefore, a large shrinkage from prior to posterior increases both the error and the computation time, while the error can be reduced by using more live points, leading to an increase in computation time. Accordingly, if the inference described here reduces the error by a factor of $\epsilon$, then we could assume that the same result can be achieved using just the standard \ac{NS} run with more live points. In particular, the number of live points would have to be increased by $\epsilon^2$, which would also increase the computational complexity by a factor of $\epsilon^2$, i.e. $T_\epsilon = \epsilon^2 T$. In contrast, the post processing with \ac{IFT} adds a constant, live point independent computational complexity $T_\text{IFT}$ to the original computational effort $T$ in order to reduce the error by $\epsilon$, i.e. $T_{\epsilon, \text{IFT}} = T + T_{\text{IFT}}$. Therefore, in cases where $T$ is high, e.g. when the \ac{KL} is high, or in particular when the evaluation of the likelihood or the sampling from the restricted prior requires a lot of  time, it makes sense to consider the inference of the likelihood instead of the addition of further live points.\\
Looking at the results in section \ref{sec:Results}, we see that the reconstruction works quite well during validation 
for the Gaussian and spike-and-slab examples. However, it is noticeable that the benefit of using post-processing decreases as the number of live points increases. This is expected because the likelihood-prior-volume function generated by \ac{NS} itself becomes smoother as the number of live points increases. Nevertheless, we can see an increase in accuracy of result for the log-evidence, up to a hundred live points, in terms of a decreased standard deviation. When the number of live points increases even further we find that the final $\sigma_\delta$, defined by the minimum distance between two adjacent nested likelihood contours, becomes very small. This can especially be seen i the case of the spike and slab likelihood for thousand livepoints, where $\sigma_\delta = 2.26 \times 10^{-10}$. As a result, the reconstruction using the Gaussian approximation of the $\delta$-function becomes numerically unstable, in terms of a high reduced $\chi^2$ value between the data and reconstruction. Therefore, the likelihood-prior-volume function inference proposed here is not applicable to a large number of live points. Instead, it could be used to improve the log-evidence calculation in scenarios where only a small number of live points are feasible, e.g. due to high \ac{LRPS} costs. The applicability, with respect to the reduced $\chi^2$ and the inherent uncertainty given by the algorithm, should therefore be checked by the user.\\
Thus, this post-processing method leaves some room for future work. First of all, it would be desirable to apply it to a problem with a large number of live points. Two different approaches could be considered. First, one could split the dataset into parts and thus generate \ac{NS} datasets with a smaller number of live points and thus a larger distance between adjacent nested likelihood contours (\citep{skilling_nested_2006}). Second, one could think of ways to avoid the Gaussian approximation of the $\delta$ function. One alternative approach that does not require the Gaussian approximation of the $\delta$ function is described in the appendix \ref{app:exact approach}. Using this approach for \ac{VI} is left for future work. Moreover, one could think of further assumptions besides the smoothness assumption and include them in the inference to allow for a wider range of likelihoods, such as non-smooth ones, or a specific set of likelihoods.  
\\
In conclusion, we have presented a post-processing step for \ac{NS} that uses a smoothness assumption to infer the likelihood-prior-volume function, providing an estimate of the inherent uncertainty in the reconstruction due to the stochastic approach. This aims to reduce the statistical error in the evidence computation introduced by the unknown prior volumes at each iteration step. Since this post-processing can deal with a varying number of live points, it is applicable to advanced \ac{NS} methods such as dynamic \ac{NS}. Finally, some work needs to be done to apply it to problems with a larger number of live points. However, the method presented here should provide a first approach to improving prior volume estimation, which remains one of the main challenges in \ac{NS}. \\
\newline
\authorcontributions{Conceptualization, M.W., J.R., P.F., W.H. and T.E.; methodology, M.W., J.R., P.F., W.H. and T.E.; software, M.W. and J.R.; validation, M.W., J.R., P.F. and T.E.; formal analysis, M.W., J.R., P.F., W.H. and T.E.; investigation, M.W., J.R., P.F., W.H. and T.E.; resources, M.W., J.R., P.F., W.H. and T.E.; data curation, M.W., W.H. and T.E.; writing---original draft preparation, M.W.; writing---review and editing, J.R., P.F., W.H. and T.E.; visualization, M.W.; All authors have read and agreed to the published version of the manuscript.}

\funding{Margret Westerkamp acknowledges support for this research through the project Universal
Bayesian Imaging Kit (UBIK, Förderkennzeichen 50OO2103) funded by the Deutsches Zentrum für
Luft- und Raumfahrt e.V. (DLR). Jakob Roth acknowledges financial support by the German Federal
Ministry of Education and Research (BMBF) under grant 05A20W01 (Verbundprojekt D-MeerKAT) and grant 05A23W01 (Verbundprojekt D-MeerKAT III).
Philipp Frank acknowledges funding through the German Federal Ministry of Education and Research for the project ErUM-IFT: Informationsfeldtheorie für Experimente an Großforschungsanlagen
(Förderkennzeichen: 05D23EO1). This research was supported by the Munich Institute for Astro-,
Particle and BioPhysics (MIAPbP) which is funded by the Deutsche Forschungsgemeinschaft (DFG,
German Research Foundation) under Germany’s Excellence Strategy- EXC-2094-390783311.}

\institutionalreview{Not applicable.}

\informedconsent{Not applicable.}

\dataavailability{Data sharing not applicable. The corresponding implementation is public and available at \url{https://gitlab.mpcdf.mpg.de/ift/public/iftns}}

\conflictsofinterest{The authors declare no conflicts of interest.} 



\abbreviations{Abbreviations}{
The following abbreviations are used in this manuscript:\\

\noindent 
\begin{tabular}{@{}ll}
geoVI & geometric variational inference\\
IFT & information field theory\\
KL & Kullback-Leibler divergence\\
LRPS & likelihood restricted prior sampling\\
MCMC & Markov Chain Monte Carlo\\
NIFTy & Numerical Information Field TheorY  \\
NS & nested sampling\\
NS-SMC & nested sampling via sequential Monte Carlo\\
VI & variational inference\\ 
\end{tabular}

}
\appendixtitles{no} 
\appendixstart
\appendix

\section[\appendixname~\thesection]{Reparametrisation}
\label{app:Reparametrization}
The aim of the reparametrisation is to map the likelihood-prior-volume function such that the correlated field describing the log-normal process is constant in the Gaussian case and non-Gaussianity is modelled by structures in the correlated field (section \ref{sec:IFT}). Below we show that the reparametrisation in eq.~\eqref{eq:repGaussianProc} leads to this condition. For the Gaussian case (eq.\eqref{eq:simpleGaussianLikelihood}) we have 
\begin{align}
\frac{d \ln L}{d \ln X} = \frac{d}{d \ln X} \biggl( - \frac{X^{2/C}}{2 \sigma^2}\biggr) = \frac{2}{C} \ln L.
\end{align}
For the Gaussian case know $\ln L_\text{max} = 0$ analytically. If we plug this in eq.~\eqref{eq:repGaussianProc} we find the desired condition, 
\begin{align}
\frac{df(\ln L)}{d \ln X} = -\frac{2}{C} = -e^{-\tau(\ln X)}, \label{eq:OffsetMean}
\end{align}
leading to $\tau(\ln X) = \text{const}$. It can be seen that the corresponding reparameterisation is given by the following equation 
\begin{align}
f(\ln L) = -\ln\biggl(\ln \frac{ L_\text{max}}{L}\biggr).
\end{align}

\section[\appendixname~\thesection]{Maximal Likelihood Calculation}
\label{app:Maximal Likelihood Calculation}
We calculate the maximum likelihood given the data on the likelihood dead contours $\vec{d}_L$. In \citep{handley_quantifying_2019} the maximum of the Shannon entropy $\mathcal{I}_\text{max} = \text{max}\ln \biggl( \frac{\mathcal{P}(\theta|d)}{\mathcal{P}(\theta)}\biggr)$ is calculated via, 
\begin{align}
\mathcal{I}_\text{max} = \mathcal{D}_\text{KL} + \frac{C}{2},
\end{align}
where $C$ is the dimension of the problem. This gives the maximum log-likelihood $\ln L_\text{max}$,
\begin{align}
\ln \biggl(\frac{L_\text{max}}{Z} \biggr) &= \int d\theta \mathcal{P}(\theta|d) \ln\biggl( \frac{L}{Z}\biggr) + \frac{C}{2} \\
\to \ln L_\text{max} &= \langle \ln L \rangle_{\mathcal{P}(\theta|d)} + C/2.
\end{align}
This can be approximated using the data on the likelihood contours and taking the deterministic \ac{NS} information on the prior volumes $\{d_{X,i}\}$ according to eq.~\eqref{eq:MeanPriorVolume},
\begin{align}
\ln L_\text{max} &= \int d\theta \frac{\mathcal{P}(d|\theta)\mathcal{P}(\theta)}{Z} \ln L + \frac{C}{2} \\
& \approx \sum_{i=1}^{n_\text{iter}}\frac{1}{2} \frac{d_{L,i} (d_{X,{i-1}} - d_{X,{i+1}})}{Z_d} \ln d_{L, i} +  \frac{C}{2}.
\end{align} 
Here, $Z_d$ is the evidence calculated according to eq.~\eqref{eq:evidence}, with $\{d_{L,i}\}$ as the likelihood information and $\{d_{X,i}\}$ as the prior volume information.

\section[\appendixname~\thesection]{Inference with Gaussian Approximation}
\label{app:exact approach}
In the following we want to work out an exact inference algorithm that is not based on a Gaussian approximation of the $\delta$ function. This approach has its difficulties due to the calculation of the Fisher metric, as we will see below, and therefore cannot be used for geoVI. The approach discussed in the main part of this paper focuses on the joint reconstruction of the likelihood-prior-volume function and the prior volumes. The approach presented here solely reconstructs the likelihood-prior-volume function, given the data on the likelihood contours. The corresponding prior volumes for each likelihood contour can then be computed by inversion of this function. Again, we incorporate the smoothness assumption into the inference, but this time using a different relationship between the likelihood-prior-volume function and the correlated field,
\begin{align}
\frac{df(\ln L)}{d \ln X} = -e^{\tau(f(\ln L))}.
\end{align}
The solution for the prior volumes given the likelihood information is, 
\begin{align}
X(f(\ln L)| \tau) = \exp \biggl( \ln X_0 - \int_{f(\ln L_0)}^{f(\ln L)} e^{-\tau(y)} dy \biggr) = \exp \biggl(- \int_{f(\ln L_0)}^{f(\ln L)} e^{-\tau(y)} dy \biggr), \label{eq:lnXDE}
\end{align}
with $X_0 = 1$.
The prior model is then given by $\mathcal{P}(\tau$), and our goal is to determine the likelihood model, denoted by $\mathcal{P}(f(\ln \vec{d}_{L})|\tau)$, which can be rewritten by marginalising over $\vec{t}$,
\begin{align}
\mathcal{P}(f(\ln \vec{d}_{L})|\tau) = \prod_{i=1}^{n_\text{iter}} \int dt_i ~ \mathcal{P}(f(d_{L,i}), \vec{t}|\tau) = 
\prod_{i=1}^{n_\text{iter}} \int dt_i ~ \mathcal{P}(t_i) ~ \mathcal{P}(f(\ln d_{L,i}) , \vec{t}|\tau). \label{eq:NonGausslklhd}
\end{align}
Here, the contraction factors $\vec{t}$ are beta distributed, i.e. $\mathcal{P}(t_i) = \text{Beta}(t_i|1,n_\text{live})$. The tricky part is to rewrite $\mathcal{P}(f(d_{L,i}) , \vec{t}|\tau)$ such that we can integrate over $t_i$. Under the assumption that we can find a unique solution for $f(\ln d_{L,i})$ with $i=1,..,n_\text{iter}$ for each of the following $n_\text{iter}$ equations using eq.~\ref{eq:lnXDE},
\begin{align}
X(f(\ln d_{L,i}) | \tau) = \prod_{j=1}^i t_j ~~~ \text{for} ~ i=1, ..., n_\text{iter}
\end{align}
the distribution $\mathcal{P}(f(\ln d_{L,i})|\tau, \vec{t})$ is defined via, 
\begin{align}
\mathcal{P}(f(\ln d_{L,i})|\tau, \vec{t}) = \delta \biggl( X(f(\ln d_{L,i}) | \tau) - \prod_{j=1}^i t_j \biggr) ~ \bigg\vert \frac{\partial X (f(\ln L)| \tau)}{\partial f(\ln L )}\bigg\vert_{f(\ln L) = f(\ln d_{L,i})}. \label{eq:interm1}
\end{align}
In the following we will denote $ |\partial_\tau X|_{d_{L,i}} = \bigg\vert \frac{\partial X (f(\ln L)| \tau)}{\partial f(\ln L )}\bigg\vert_{f(\ln L) = f(\ln d_{L,i})}$ and $X_i = X(f(\ln d_{L,i}) | \tau)$, for the matter of brevity of the equations. With the definition of $t_0:=1$, we can further rewrite eq.~\eqref{eq:interm1} to,
\begin{align}
\mathcal{P}(f(\ln d_{L,i})|\tau, \vec{t}) = \delta \biggl( t_i - \frac{X_i}{\prod_{j=0}^{i-1} t_j }\biggr) ~ |\partial_\tau X|_{d_{L,i}} ~ \prod_{j=0}^{i-1} \frac{1}{t_j} = \delta \biggl( t_i - \frac{X_i}{X_{i-1}}\biggr) ~ |\partial_\tau X|_{d_{L,i}} ~ \prod_{j=0}^{i-1} \frac{1}{t_j}
\end{align} 
Using this and eq. \ref{eq:NonGausslklhd}, we can write down the overall reconstruction likelihood for this approach:
\begin{align}
\mathcal{P}(f(\ln \vec{d}_L)|\tau) &= \prod_{i=1}^{n_\text{iter}} \int d t_i \mathcal{P}(t_i) \delta \biggl( t_i - \frac{X_i}{X_{i-1}}\biggr) ~ |\partial_\tau X|_{d_{L,i}} ~ \prod_{j=0}^{i-1} \frac{1}{t_j} \\
 &= \prod_{i=1}^{n_\text{iter}} \text{Beta}\biggl(\frac{X_i}{X_{i-1}}| 1, n_\text{live} \biggr) \frac{1}{X_{i-1}}  \bigg |\partial_\tau X|_{d_{L,i}}\\
 &= \prod_{i=1}^{n_\text{iter}} n_{\text{live}, i} \frac{X_i^{n_{\text{live}, i}-1}}{X_{i-1}^{n_{\text{live}, i}}}~|\partial_\tau X|_{d_{L,i}}
\end{align}
So far, the likelihood model and the prior model can be used for a maximum a posteriori approximation of $f(\ln L)$. For variational inference (VI), however, the Fisher metric is required,
\begin{align}
(I_F)_{k, l} = \int d f(\ln d_{L,0}) ... \int  d f(\ln d_{L,n_\text{iter}}) \frac{\partial}{\tau_k} (-\ln(\mathcal{P}(f(\ln d_L)|\tau))) \frac{\partial}{\tau_l} (-\ln(\mathcal{P}(f(\ln d_L)|\tau)))
\end{align} 
Due to the complexity and length of the calculations involved in this integration and the difficulties of its implementation, we have opted to use the Gaussian approximation and reserve this approach for future work.
\begin{adjustwidth}{-\extralength}{0cm}

\reftitle{References}



\bibliography{bib}

\PublishersNote{}
\end{adjustwidth}
\end{document}